\documentclass[twocolumn]{aa} 
\usepackage[switch,modulo]{lineno}

\usepackage{graphicx}
\usepackage{txfonts}
\usepackage{natbib}
\bibpunct{(}{)}{;}{a}{}{,} 

\newcommand{\dnumoy}{\langle\Delta\nu\rangle}

\newcommand{\logg}{$\log g$}
\newcommand{\feh}{[Fe/H]}

\newcommand{\vsini}{$v \sin i$}

\newcommand{\ind}[1]{_{\mathrm{#1}}}

\newcommand{\teff}{$T_{\rm eff}$}
\newcommand{\vmicro}{$v_{\rm micro}$}
\newcommand{\numax}{\nu\ind{max}}

\def\numax{\nu\ind{max}}
\def\nmax{n\ind{max}}
\newcommand\Tg{\Delta\Pi_1}

\newcommand{\np}{n\ind{p}}
\def\Dnu{\Delta\nu}

\def\ng{n\ind{g}}
\def\offsetg{\varepsilon\ind{g}}
\def\offsetp{\varepsilon\ind{p}}

\begin{document}

 \title{Study of HD~169392A observed by CoRoT \thanks{The CoRoT space mission, launched on December 27th 2006, has been developed and is operated by CNES, with the contribution of Austria, Belgium, Brazil, ESA (RSSD and Science Programme), Germany and  Spain.} and HARPS\thanks{This work is based on ground--based observations made with the ESO 3.6m-telescope at La Silla Observatory under the ESO Large Programme  LP185-D.0056.}}

 \titlerunning{Study of HD~169392A}

   \author{S. Mathur\inst{1, 2} \and 
                 H. Bruntt\inst{3} \and
                 C. Catala\inst{4}\and
                 O. Benomar\inst{5} \and   
                 G.~R. Davies \inst{2} \and
                 R.~A. Garc\'\i a\inst{2} \and
                 D. Salabert\inst{6} \and                
                 J. Ballot\inst{7,8} \and
                  B. Mosser\inst{4} \and
                  C. R\'egulo\inst{9,10} \and
                 W.~J. Chaplin\inst{11} \and     
                 Y. Elsworth\inst{11} \and
                 R. Handberg\inst{3}\and
                 S. Hekker\inst{12, 11} \and
                 L. Mantegazza\inst{13} \and
                 E. Michel\inst{4}\and
                 E. Poretti\inst{13}\and
                 M. Rainer\inst{13} \and
	       I.~W. Roxburgh\inst{14} \and
	       R. Samadi\inst{4}\and
	       M. St\c e\' slicki\inst{15,1}
	       K. Uytterhoeven\inst{9,10} \and
	       G.~A. Verner\inst{11} \and
	       M. Auvergne\inst{4} \and
	       A. Baglin\inst{4}\and
	        S. Barcel\'o Forteza\inst{9,10} \and
	       F. Baudin\inst{16} \and
	       T. Roca Cort\'es\inst{9,10} 
	}
   \offprints{savita.mathur@gmail.com}
   \institute{High Altitude Observatory, NCAR, P.O. Box 3000, Boulder, CO 80307, USA
   \and Laboratoire AIM, CEA/DSM -- CNRS - Univ. Paris Diderot -- IRFU/SAp, Centre de Saclay, 91191 Gif-sur-Yvette Cedex, France
   \and Danish AsteroSeismology Centre, Department of Physics and Astronomy, University of Aarhus, 8000 Aarhus C, Denmark
      \and LESIA, UMR8109, Universit\'e Pierre et Marie Curie, Universit\'e Denis Diderot, Obs. de Paris, 92195 Meudon Cedex, France
   \and Sydney Institute for Astronomy, School of Physics, University of Sydney, NSW 2006, Australia
   \and Laboratoire Lagrange, UMR7293, Universi\'e de Nice Sophia-Antipolis, CNRS, Observatoire de la C\^ote d'Azur, 06304 Nice Cedex 4, France
  \and CNRS, Institut de Recherche en Astrophysique et Plan\'etologie, 14 avenue Edouard Belin, 31400 Toulouse, France
  \and Universit\'e de Toulouse, UPS-OMP, IRAP, 31400 Toulouse, France 
     \and Instituto de Astrof\'\i sica de Canarias, 38205, La Laguna, Tenerife, Spain
        \and Universidad de La Laguna, Dpto de Astrof\'isica, 38206, Tenerife, Spain
    \and School of Physics and Astronomy, University of Birmingham, Edgbaston, Birmingham B15 2TT, UK  
   \and Astronomical Institute Anton Pannekoek, U. of Amsterdam, PO Box 94249, 1090 GE Amsterdam, NL
   \and INAF - Osservatorio Astronomico di Brera, via E. Bianchi 46, 23807, Merate (LC), Italy
   \and Astronomy Unit, Queen Mary University of London, Mile End Road, London E1 4NS, UK
  \and Instytut Astronomiczny, Uniwersytet Wroclawski, Kopernika 11, 51-622, Wroclaw, Poland
   \and Institut d'Astrophysique Spatiale, UMR8617, Universit\'e Paris XI, Batiment 121, 91405 Orsay Cedex, France
   }

   \date{Received 2012; accepted }

 \abstract
     {The numerous results obtained with asteroseismology thanks to space missions such as CoRoT and \emph{Kepler} are providing a new insight on stellar evolution. After five years of observations, CoRoT is going on providing high-quality data. We present here the analysis of the  double star HD~169392 complemented by ground-based spectroscopic observations.}
   {This work aims at characterizing the fundamental parameters of the two stars, their chemical composition, the acoustic-mode global parameters including their individual frequencies, and their dynamics.}
   {We have analysed HARPS observations of the two stars to retrieve their chemical compositions. Several methods have been used and compared to measure the global properties of acoustic modes and their individual frequencies from the photometric data of CoRoT. }
   { The new spectroscopic observations and archival astrometric values
suggest that HD 169392 is a wide binary system weakly bounded.
We have obtained the spectroscopic parameters for both components, suggesting the origin from the same cloud. However, only the mode signature of HD~169392 A has been measured within the CoRoT data. The signal-to-noise ratio of the modes in HD~169392B is too low to allow any confident detection. We were able to extract mode parameters of modes for  $\ell$=0, 1, 2, and 3. The study of the splittings and inclination angle gives two possible solutions with splittings and inclination angles of 0.4-1.0\,$\mu$Hz and 20-40$\degr$ for one case and 0.2-0.5\,$\mu$Hz and 55-86$\degr$ for the other case. The modeling of this star with the Asteroseismic Modeling Portal led to a mass of 1.15\,$\pm\,0.01$\,M$_{\odot}$, a radius of 1.88\,$\pm\,0.02\,$R$_{\odot}$, and an age of 4.33\,$\pm\,0.12$\,Gyr, where the uncertainties are the internal ones.}
  {}

   \maketitle
   
\section{Introduction}

The convective motions in the external layers of solar-like oscillating stars excite acoustic sound waves, which become trapped in the stellar interiors \citep[e.g.,][for a detailed review]{1977ApJ...212..243G,Samadi11}. The precise frequencies of these pressure (p) driven waves depend on the properties of the medium in which they propagate. Thus, stellar seismology allows us to infer the properties of the Sun and stellar interiors by studying and characterizing these p modes \citep[e.g.,][]{GouKos1996,JCD2002}. However, our knowledge of the properties of the stellar interiors depends on our ability to correctly measure the oscillations modes \citep[e.g.][]{AppGiz1998,2011arXiv1103.5352A}.

In recent years the French-led satellite, Convection, Rotation and planetary Transits \citep[CoRoT,][]{2006cosp...36.3749B}, and NASA's {\it Kepler} mission \citep{2010ApJ...713L..79K,2010Sci...327..977B} have been providing high-quality, long-term seismic data of solar-like stars. CoRoT has studied some interesting F- and G-type stars \citep[e.g.][]{2008A&A...488..705A,2009A&A...506...51B,2009A&A...506...41G,2010A&A...515A..87D}  including some hosting planets \citep{2010A&A...518L.153G,2011A&A...530A..97B,2012ApJ...746..123H} and others in binary systems  \citep{2010A&A...518A..53M}. {\it Kepler} observations have enabled ensemble asteroseismology of hundreds of solar-like stars \citep[e.g.][]{2011Sci...332..213C} as well  as precise studies of long time series with more than 8 months of nearly continuous data \citep{2011A&A...534A...6C,2011ApJ...733...95M,2012arXiv1204.3147A}, and pulsating stars in binary systems \citep[e.g.][White et al. in prep.]{2010ApJ...713L.187H} and in clusters \citep[e.g.][]{2011ApJ...729L..10B, 2011ApJ...739...13S}. The fundamental properties of stars can be inferred by means of asteroseismic quantities either by using scaling relations \citep[e.g.][]{1995A&A...293...87K,2011A&A...529L...8K,2007A&A...463..297S},  or by using stellar models \citep[e.g.][]{2009A&A...506..175P,2010ApJ...723.1583M,2012ApJ...748L..10M,2012A&A...537A.111C,2012A&A...543A..96E,2012ApJ...749..152M,2012A&A...540A..31M}. This allows us to have some information on the structure of the stars but also to test the models and start improving the physics included in stellar evolution codes \citep[e.g.][and references there in]{2011arXiv1104.5191C}.


In the present paper we report the study of the double star HD~169392, using high-resolution spectrophotometry by HARPS combined with seismic analysis from CoRoT photometry. However in this latter case, only  HD~169392A was analyzed because the 
HD 169392B was too faint. The two components
of HD 169392$\equiv$WDS 18247-0636 have a separation of 5.881$\pm$0.006\,\arcsec, a position angle of 195.08$\pm$0.09 \degr, and a magnitude
difference $\Delta V$=1.507$\pm$0.002 at the epoch 2005.48
\citep{2007A&A...472.1055S}. The two components are reported in SIMBAD{\footnote {This research has made use of the SIMBAD database, operated at CDS, Strasbourg, France.}} as
having similar proper motions. HD 169392A$\equiv$HIP 90239 is the main
component of the pair, with spectral type G0~IV and a parallax $\pi=13.82\,\pm\,1.20$\,mas \citep{2007A&A...474..653V}. From the Geneva-Copenhagen survey 
\citep{2004A&A...418..989N,2007A&A...475..519H} this primary star has $T_{\rm eff}$ = 5942 K, [Fe/H] = -0.03, $m_V$ = Ê7.50 mag, and $v\sin i$ = 3 $\rm{km\,s^{-1}}$. The second component of the system, HD~169392B \citep{1984AJ.....89..515D} is a G0 V - G2 IV star with $m_V$ = 8.98 mag.

New spectroscopic observations performed by HARPS \citep{2003Msngr.114...20M} and detailed in Section~\ref{sec:spectro} 
have allowed a precise characterization of the stellar properties of the  two stars in the system. In Section 3, we describe the methodology used to prepare the original time series from CoRoT, to fit the mode oscillation power spectrum, and to derive a single frequency set from results of different fitters. We derive the global properties of the modes of both stars (Section 4) while in Section 5 we analyse the surface rotation and the background of HD~169392A. In Section 6, we present the peak-bagging methods used to obtain the frequencies of the individual p modes of HD~169392A and the results are discussed in Section 7. With the spectroscopic constraints derived and the frequencies of modes, we modeled the star and retrieve the fundamental parameters of HD~169392A in Section 8. Finally,  the main results are summarized in Section 9.

\section{Fundamental stellar parameters}
\label{sec:spectro}


High-quality spectra of HD 169392 were obtained in the
framework of the CoRoT Ground-Based Observations Working Group \citep[e.g.][]{2006ESASP1306..329C,2008JPhCS.118a2077U,2012arXiv1202.3542P}. We obtained high signal-to-noise ratio (S/N) spectra
of each component using HARPS in the high-resolution
mode ($R$=114,000). The spectrum of HD 169392A taken
on the 25-26 June 2011 night shows a radial velocity (RV) of
$-68.1606\pm0.0004$ km\,s$^{-1}$, that of HD169392B taken
on the 24-25 June 2011 shows RV=$-70.1388\pm0.0004$ km\,s$^{-1}$.  To have
common proper motions and compatible radial velocities is
a sufficient condition
to consider a double star as a physical pair \citep[e.g.][]{1955ApJ...121..670S}.
In the case of HD 169392, the separation of 0.0588
\arcsec at a distance of 723 pc corresponds to 4250 AU and then
the system should be  weakly gravitationally bound.

The HARPS spectra have both an excellent
S/N, i.e., S/N=267 for the A component, and S/N=234 for the B one. They were
reduced using a MIDAS semi-automatic pipeline \citet{RainerPhD2003}. Their high
quality  allowed a careful and detailed analysis by means of the
semi-automatic pipeline VWA \citep{2010MNRAS.405.1907B}. More than 400 spectral lines have been analyzed in each star. The match of the synthetic line profile to the observations was inspected by eye and several lines were rejected either due to problems with the placement of the continuum
or due to strong blends in the line wings. Atmospheric models were determined by interpolating on a grid of MARCS models \citep{gustafsson08} and line data were extracted from VALD \citep{kupka99}. 
The oscillator strengths ($\log gf$) were adjusted relative to a solar spectrum as described by \citet{2012arXiv1203.0611B}.

The classical analysis method consists of treating the atmospheric parameters of the effective temperature, \teff, the surface gravity, \logg, and microturbulence, $v_{\rm macro}$ as free parameters. The optimum solution is then selected to minimize the correlation between Fe\,{\sc i} abundance, equivalent width, and excitation potential. Furthermore, we required that the mean abundance determined from neutral and ionized Fe lines agree. The final parameters are listed in Table~\ref{tab:param}. For these parameters we calculated the abundances of 13 different
species. We have done the analysis with and without the seismic \logg, which is available only for the A component (see Section 4.1).  When the seismic value is used, it affects the \logg\ value by $+$0.2\,dex and \teff\ by $+$100\,K. We also note that \feh\ is increased by about 0.05 dex.

\begin{figure}[!htbp]
\begin{center}
\includegraphics[width=9cm, trim=4cm 2cm 4cm 14cm]{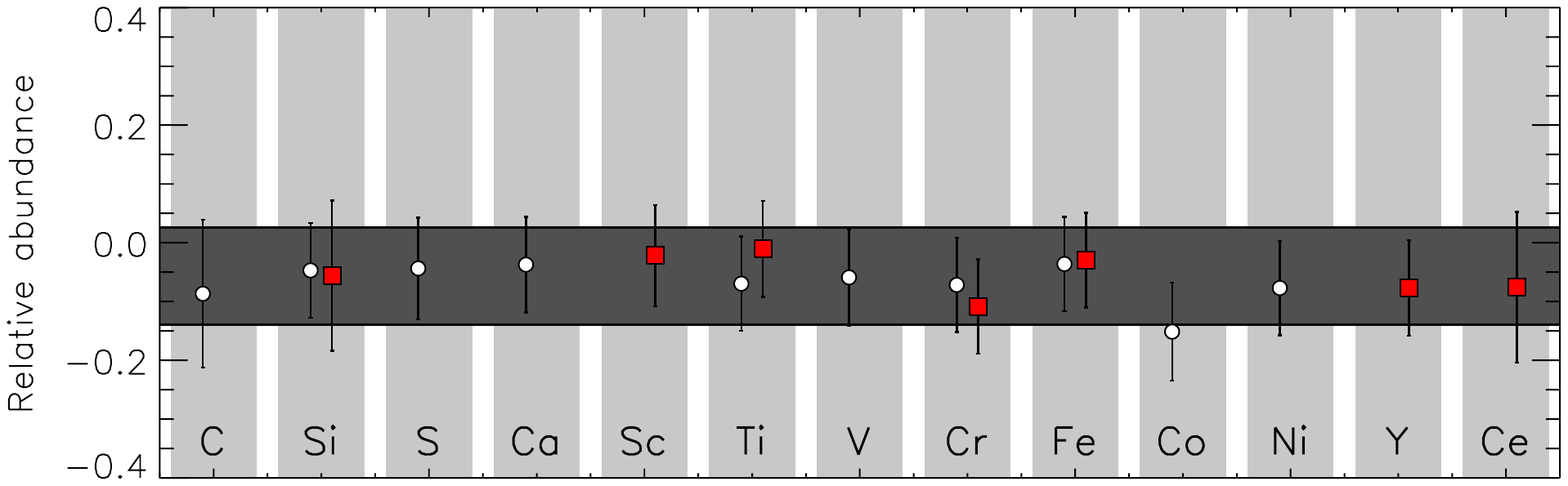}\\
\includegraphics[width=9cm, trim=4cm 1cm 4cm 13cm]{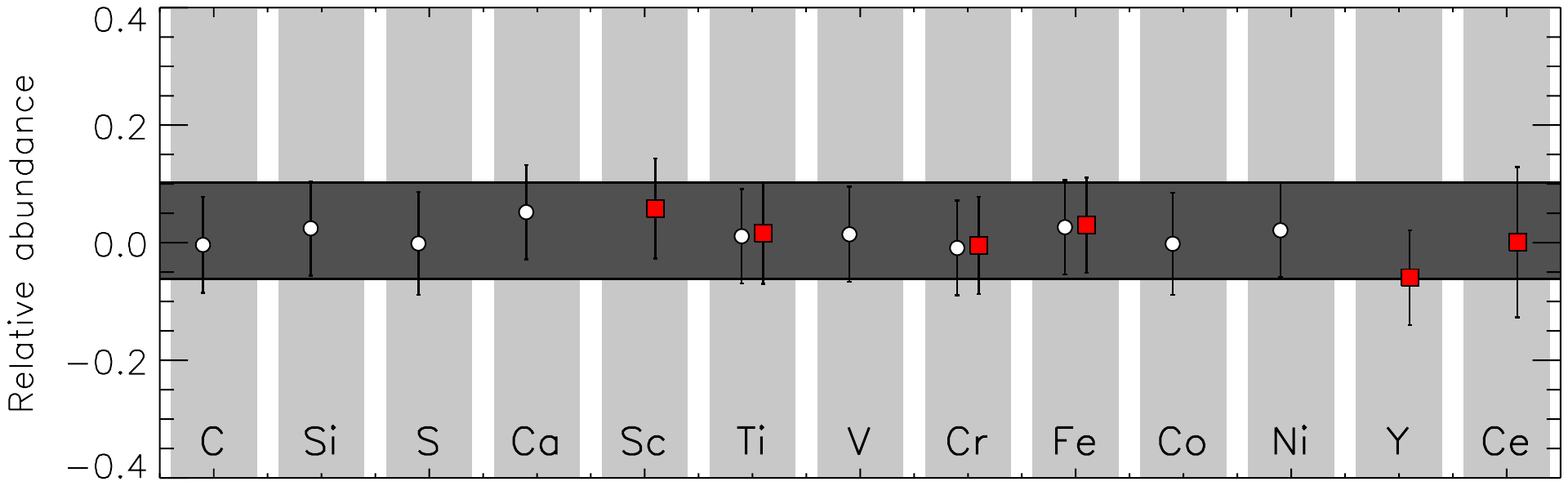}
\caption{Abundances for 14 elements relative to the Sun for the A (top) and B (components) of the system. Circles and box symbols are used for the mean abundance from neutral and singly ionized lines, respectively. The dark grey horizontal bar marks the mean metallicity within 1-$\sigma$ uncertainty range. The lithium is not plotted because it has a much higher value than the other elements in the A component.}
\label{Fig:comp}
\end{center}
\end{figure}

We listed the abundances relative to the Sun ($\Delta$A) along with the number of spectral lines used in Table~\ref{tab:abund} and plotted them in Fig.~\ref{Fig:comp}. The lithium abundance has been determined for both components of the system:  $\rm{[Li/H]_A}=+1.42 \pm 0.10$ and $\rm{[Li/H]_B}=+0.23 \pm 0.12$ (see details in \citet{2012arXiv1203.0611B}). According to the temperatures of both stars, the measured lithium abundances are consistent with the sample analyzed for example by \citet{2004A&A...414..601I}. 

We have adjusted the projected splittings, \vsini\ value by fitting the observations
to synthetic line profiles for several isolated lines. During these fits the macroturbulence was fixed using the 
calibration for solar-type stars from \citet{2010MNRAS.405.1907B}. We note that the metallicities of the two components are quite similar.

\begin{table}
 \centering
 \caption{Table with spectroscopic parameters for both A  and B components of the system derived from our HARPS observations and the seismic \logg.
\label{tab:param}}
\begin{tabular}{c|ccc}
\hline
\hline
        &  A      &       B   &         Units \\
\hline

$T_{\rm eff}$     &  5985 $\pm$  60   & 5885 $\pm$ 70   & K\\
\logg     &  3.96 $\pm$ 0.07  & 3.75 $\pm$0.07  & cgs\\
\feh     & $-$0.04 $\pm$ 0.10 & $-$0.10 $\pm$ 0.10& dex \\
\vmicro   & 1.41$\pm$0.07  & 1.04 $\pm$ 0.07 & km~s$^{-1}$\\
\vsini     & 1.0  $\pm$ 0.6   & 3.0 $\pm$ 0.7   & km~s$^{-1}$\\
$v_{\rm macro}$   & 2.8  $\pm$ 0.4   & 2.5 $\pm$0.4  & km~s$^{-1}$ (fixed)\\
\hline
\end{tabular}
\end{table}

\begin{table}
 \centering
 \caption{Abundances relative to the Sun ($\Delta A$) and the number of lines ($N$) used
 in the spectral analysis of the two components (A and B) of the binary system HD~169392.The abundances have an uncertainty of 0.08\,dex for all the elements excepting for the lithium, which are 0.10 and 0.12 for A and B components respectively.
\label{tab:abund}}
\begin{tabular}{l|lr|lr}
\hline
\hline
   Element & $\Delta A$(A) (dex) & $N$ (A)  & $\Delta A$(B)  (dex)& $N$ (B) \\
\hline
 {\bf  {Li  \sc   i}} &     $ +1.42 $    &   1  &  $  +0.23 $   &   1    \\      
  {C  \sc   i} &     $ -0.09 $    &   4  &  $ -0.00 $   &   3    \\      
  {Si \sc   i} &     $ -0.05 $    &  25  &  $ +0.02 $   &  29    \\      
  {Si \sc  ii} &     $ -0.06 $    &   1  &              &        \\
  {S  \sc   i} &     $ -0.04 $    &   3  &  $ -0.00 $   &   3   \\      
  {Ca \sc   i} &     $ -0.04 $    &  13  &  $ +0.05 $   &  10   \\      
  {Sc \sc  ii} &     $ -0.02 $    &   5  &  $ +0.06 $   &   4   \\      
  {Ti \sc   i} &     $ -0.07 $    &  28  &  $ +0.01 $   &  43  \\      
  {Ti \sc  ii} &     $ -0.01 $    &  12  &  $ +0.02 $   &   8   \\      
  {V  \sc   i} &     $ -0.06 $    &   8  &  $ +0.01 $   &  11   \\      
  {Cr \sc   i} &     $ -0.07 $    &  22  &  $ -0.01 $   &  23   \\      
  {Cr \sc  ii} &     $ -0.11 $    &   4  &  $ -0.00 $   &   6   \\      
  {Fe \sc   i} &     $ -0.06 $    & 239  &  $ +0.03 $   & 233   \\      
  {Fe \sc  ii} &     $ -0.03 $    &  22  &  $ +0.03 $   &  20   \\      
  {Co \sc   i} &     $ -0.15 $    &   6  &  $ -0.00 $   &   7   \\      
  {Ni \sc   i} &     $ -0.08 $    &  64  &  $ +0.02 $   &  75   \\      
  {Y  \sc  ii} &     $ -0.08 $    &   4  &  $ -0.06 $   &   3   \\      
  {Ce \sc  ii} &     $ -0.08 $    &   2  &  $ +0.00 $   &   2   \\      
  \hline   
\end{tabular}
\end{table}

\subsection{Deriving $\nu_{\rm max}$ from spectroscopic parameters}
\label{sect:sr}
The proportionality between the frequency at maximum power, $\nu_{\rm max}$, and the acoustic cutoff frequency was originally suggested by \citet{1991ApJ...368..599B}, developed by \citet{1995A&A...293...87K}, and justified theoretically by \citet{2011A&A...530A.142B}. According to theory, the large frequency spacing, $\Delta \nu$, is proportional to the square root of the mean density of the star. The scaling relations -- scaled to solar values-- have also been extensively tested with observational data \citep[e.g.][]{2011ApJ...743..143H}. The relations allow us to obtain approximate values for the so-called mean ``large spacing'', $\dnumoy$, and the frequency of maximum amplitude $\nu_{\rm max}$. 




Combining both equations, we are able to deduce $\nu_{\rm max}$ as a function of the effective temperature and the \logg\, derived in Table~\ref{tab:param} for both components of the binary system. 
Taking into account the uncertainties in the observations we have in the spectroscopic parameters, we obtain a range of $\nu_{\rm max}$ for HD~169392A of [524, 730] $\mu$Hz, while for HD~169392B the range is [2202, 3077] $\mu$Hz.

\section{CoRoT photometric light curve}
\label{Sect:LC}
CoRoT has observed HD~169392 (CoRoT id = 9161) continuously during 91.2 days starting on 2009 April 1 till 2009 July 2. These observations correspond to the third long run in the center direction of the galaxy (LRc03). In the present study, we have used the so called {\it helreg} level 2 (N2) datasets \citep{2007astro.ph..3354S}, i.e., time series prepared by the CoRoT Data Center (CDC) that are regularly cadenced at 32s in the heliocentric frame.

During the crossing of the South Atlantic Anomaly (SAA), the CoRoT measurements are perturbed \citep{2009A&A...506..411A} and the power spectrum shows a sequence of spikes at frequencies of $n \times 161.7$ $\mu$Hz where $n$ is an integer. Moreover, other spurious peaks appear at multiples of daily harmonics  at  ($n\,\times 161.7)\,\pm$\,($m\,\times$\,11.57$)\,\mu$Hz, where $m$ is an integer. Therefore, some teams in the collaboration used different interpolation techniques to fill the perturbed data. One method used  an ``inpainting'' algorithm --a Multi-Scale Discrete Cosine Transform \citep{2010arXiv1003.5178S}-- because we obtained good results for other CoRoT targets \citep[see for example HD~170987,][]{2010A&A...518A..53M}. The overall duty cycle before the interpolation was 83.4\%. Another method was to interpolate the gaps produced by SAA with parabola fitting the points around the gap, in a way similar to the usual interpolation performed in N2 data for missing points \citep[see][for details]{2011A&A...530A..97B}.

\begin{figure}[!htbp]
\centering
\includegraphics[width=9.5cm]{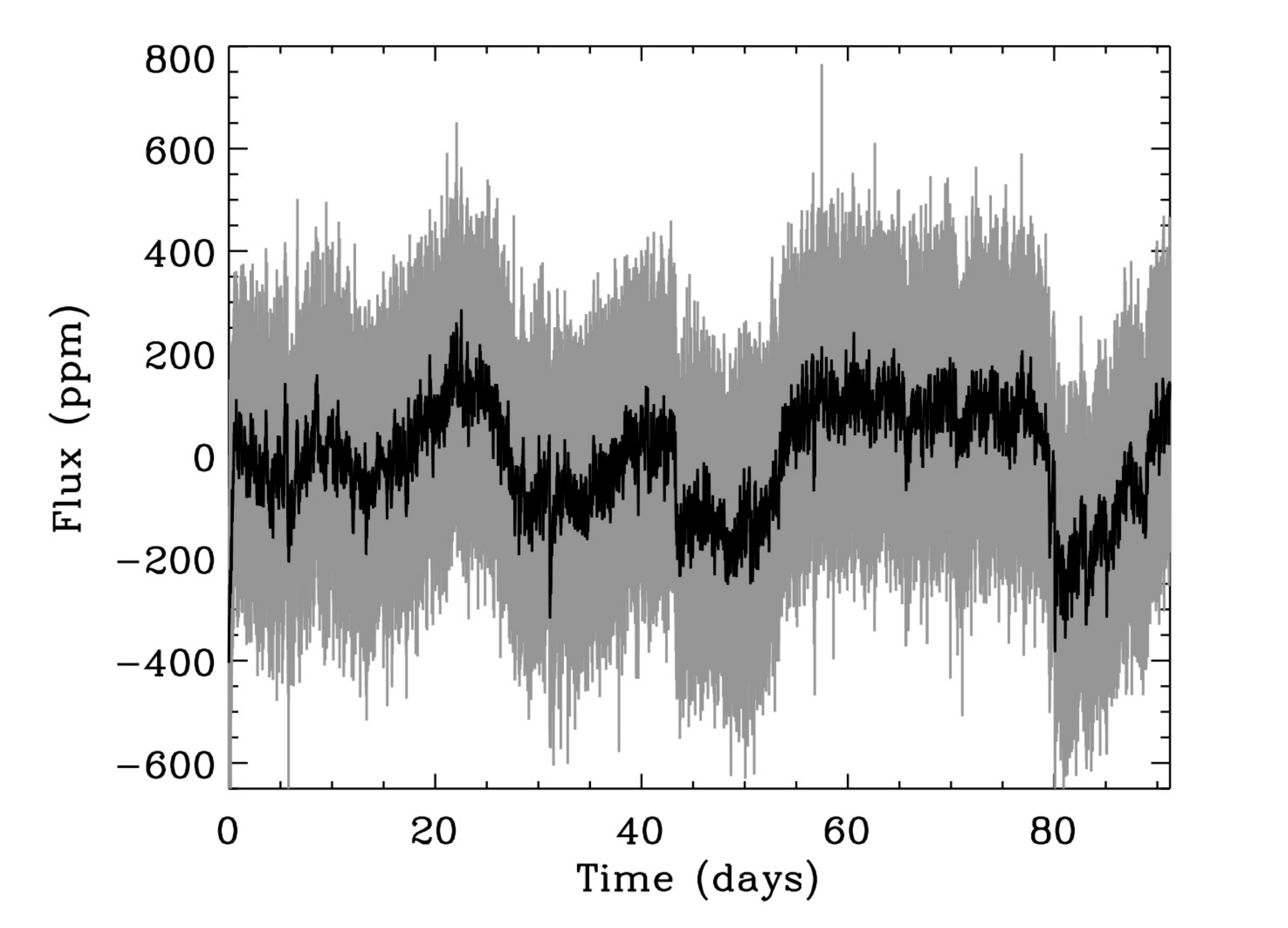}
\caption{N2-helreg relative flux (in grey) after inpainting the bad points and those  taken during the SAA crossing. The aging has been removed by a sixth order polynomial fit. The black curve corresponds to a 1 hour rebin.}
\label{Fig:flux}
\end{figure}

The measured light curve has been converted into a relative flux (ppm) by correcting first for any discontinuity in the flux and then removing a sixth order polynomial fit to take into account the aging of the instrument \citep{2009A&A...506..411A}.  The resultant flux is plotted in Fig.~\ref{Fig:flux}. Two jumps --of possible instrumental origin-- are visible at the dates 43.3 and 80.05 days. Although they have no significant influence in the power spectral density (PSD), they have an important impact on the analysis of the surface rotation (see Section~\ref{rotation}). To compute the PSD, we used a standard fast Fourier transform algorithm and normalized it as the so-called one-sided power spectral density \citep{1992nrfa.book.....P}. Finally, we calibrated the spectrum to  comply with the Parseval's theorem.

\section{Global seismic parameters of the binary system}
\label{Sect:global}
\subsection{HD~169392A}

Three different teams, A2Z, COR and OCT, analyzed the global seismic properties of HD~169392A using different methodologies. Explanation of each method can be found in  \citet{2010A&A...511A..46M}, \citet{2009A&A...508..877M} and \citet{2010MNRAS.402.2049H}, respectively. Each team estimated the mean large frequency spacing, $\dnumoy$, the frequency of maximum power in the p-mode bump, $\nu_{\rm max}$, and the maximum amplitude per radial mode, converted to the bolometric one following the method described in \citet{2009A&A...495..979M}, $A_{\rm bol,l=0}$. 
The results from A2Z are summarized in Table~\ref{Tab:global}.  Briefly, the A2Z and OCT methods analyse the power spectrum of the power spectrum to measure the mean large separation, while the COR method computes the envelope autocorrelation function. All teams show good agreement on the values of the seismic parameters when we take into account the different frequency ranges used for the calculations (a full comparison of them can be found in \citet{2011A&A...525A.131H} and \citet{2011MNRAS.tmp..892V}).
\begin{table}
 \centering
 \caption{Global seismic parameters computed using the A2Z pipeline.
\label{Tab:global}}
\begin{tabular}{c|cc}
\hline
\hline
Quantity        &  Value      &         Units \\
\hline

$\dnumoy$    &  56.32\,$\pm$ \,1.17    & $\mu$Hz\\
$\nu_{\rm max}$     &  1030\,$\pm$\,54  & $\mu$Hz\\
$A_{\rm bol,l=0}$     & 3.61\,$\pm$\,0.35 & ppm\\

\hline
\end{tabular}
\end{table}

The values of $\nu_{\rm max}$ and $A_{\rm bol,l=0}$ were obtained by smoothing the background-subtracted PSD with a sliding window of width 2$\dnumoy$ as described by  \citet{2008ApJ...682.1370K}. Then, to derive $A_{\rm bol,l=0}$  we followed the procedure to correct for the CoRoT instrumental response as described in \citet{2008Sci...322..558M}, while $\nu_{\rm max}$ corresponds to the frequency of the maximum power of a fitted Gaussian. 

Using  scaling relations from solar values \citep[e.g.][]{1995A&A...293...87K} and the effective temperature derived by our HARPS observations (see Table~\ref{tab:param}), we can infer some seismic stellar parameters. Therefore we obtain a mass of 1.34\,$\pm$\,0.26\,M$_{\sun}$, a radius of 1.97\,$\pm$\,0.19\,R$_{\sun}$ and a \logg\ of 3.96 $\pm$ 0.013 dex. 

\subsection{Looking for HD~169392B}

We derived in Section~\ref{sect:sr} that the maximum of the p-mode bump for HD~169392B should lie around a region centered at  $\sim$~2650~$\mu$Hz. Five different teams looked for the signature of this second star in a blind way, without success. We failed to obtain a possible detection at a level of 90\,$\%$ confidence level in all the cases. 
{With the stellar parameters of the two components (\teff,  $m_V$, $R$), we computed the probability of detection of the modes following the method described in \citet{2011ApJ...732...54C}. For the A component, we obtain a 100\,\% probability of detection. For the B component, which has its flux dimmed by the A component by 80\,\%, and by assuming a radius of $\sim$\,1.03\,$R_{\odot}$ and a mass of $\sim$\,1\,$M_{\odot}$ from the scaling relations, we obtain a probability of 2\,\% to detect the modes. This agrees with the non-detection with the CoRoT data.

We focus the remainder of the paper on the A component of the system.

 \section{Background and surface Rotation of HD~169392A}
 
 \subsection{Stellar background}
 \label{Sect:back}
From the measured flux we compute the PSD using a fast Fourier transform algorithm as points are equidistant in time in the light curve. The result is plotted in Fig.~\ref{Fig:back} (grey line). The PSD rebinned over 12.8\,$\mu$Hz (101 bins) and over the large separation are also shown (black and green lines respectively in Fig.~\ref{Fig:back}).

A single p-mode bump is clearly visible corresponding to HD~169392A. The convective background (red curve in Fig.~\ref{Fig:back}) can be fitted using a standard maximum likelihood estimator (MLE) with 3 standard components: a white noise (W), one Harvey law \citep{1985ESASP.235..199H}, and one power law \citep[see for more details, e.g.,][]{2011ApJ...741..119M}:

\begin{equation}
B(\nu)=W+ \frac{4 \tau_g \sigma^2_g}{1+(2\pi\tau_g\nu)^{\alpha_g}} + P_a\nu^{-e_a}
\end{equation}
where $\tau_g$ and $\sigma_g$ are the characteristic time scale and amplitude of the granulation, and $\alpha_g$ and $e_a$ are the exponents characterizing the temporal coherence of the phenomenon. The obtained coefficients are: $W=0.945 \pm 0.003$ ppm$^2/\mu$Hz, $\tau_g=557 \pm 60$ s, $\sigma_g=46.8 \pm 2.1$ ppm, $\alpha_g=2.43 \pm 0.18$, $P_a=24\,\pm\,4.3$, and $e_a = 1.8\,\pm\,0.06$. The value obtained for the granulation time scale agrees with the relation derived with the {\it Kepler} red giants with $\nu_{\rm max}$ \citep{2011ApJ...741..119M}. Finally, we also computed the rms amplitude of the granulation:

\begin{equation}
\sigma_{\mathrm{g, rms}} = \sigma_g / \sqrt{\frac{\alpha_g}{2}\sin\left(\frac{\pi}{\alpha_g}\right)}
\end{equation}

\noindent and we obtain $\sigma_{\mathrm{g,rms}} \approx$\,43.3\,$\pm$\,2.2\,ppm. Nevertheless, we have to keep in mind that this measurement is polluted by the secondary star of the binary system. There are two opposite effects. HD~169392B contributes for 20\% to the total observed flux (see also Sect.~7.6). This translates into an underestimation of $\sigma_{\mathrm{g,rms}}$ by a similar amount. On the other hand, the granulation component measured for HD~169392A probably includes a contribution from the granulation (and other short-frequency variabilities) of HD~169392B. This would lead to an overestimation of $\sigma_{\mathrm{g,rms}}$.

 \begin{figure}[!htbp]
\centering
\includegraphics[width=9cm]{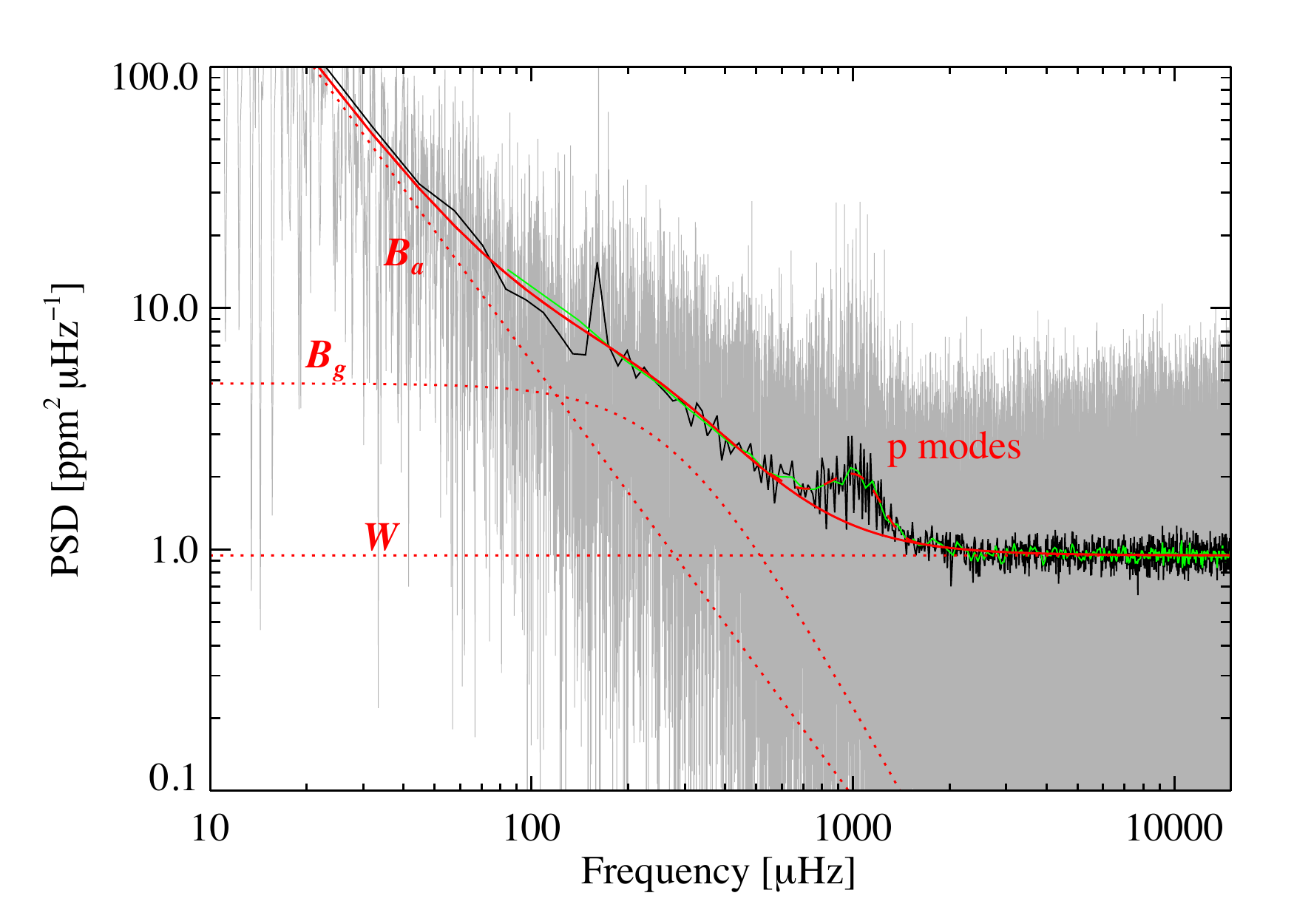}
\caption{PSD of HD~169392 data (grey) that has been modeled as explained in the text (red curves) using the usual three components: white noise (W), granulation noise ($B_g$) and stellar activity and/or large scales of convection  ($B_a$). The red continuous line is the sum of the three components above mentioned, resulting from the fit of the smoothed spectrum over $\Delta \nu$ (green line). The black curve corresponds to the spectrum smoothed over 101 bin, i.e. 12.8\,$\mu$Hz.}
\label{Fig:back}
\end{figure}

\subsection{Surface rotation}
\label{rotation}
Magnetic features such as starspots crossing the visible stellar disk of a star produce a fluctuation of the flux emitted by the star. The careful analysis of these long-period modulations in the observed light curve provides invaluable information on the average rotation rate of the surface of the star, at the latitudes where these magnetic features evolve. This study can be done directly in the light curve by modeling the spots \citep[e.g.][]{2009A&A...506...33M,2009A&A...506..245M}, or by the close inspection of the low-frequency part of the power spectrum \citep[e.g.][]{2009A&A...506...41G,2011A&A...534A...6C}.

Looking at the light curve displayed in Fig.~\ref{Fig:flux} we do not see any clear modulation produced by spots. However, the low-frequency part of the power spectrum is dominated by two significant peaks at 0.25 and 0.64 $\mu$Hz (see black line of Fig.~\ref{Fig:lowpsd}), the latter being a harmonics of the first peak when we consider the frequency resolution of 0.127 $\mu$Hz. But as explained in Section~\ref{Sect:LC}, there are two jumps of possible instrumental origin at 43.3 and 80.05 days that could be at the origin of the peak at 0.25 $\mu$Hz. To verify this hypothesis, we correct the jumps using the same procedure used to correct {\it Kepler} data \citep{2011MNRAS.414L...6G}. In the PSD of the resultant light curve (blue line in Fig.~\ref{Fig:lowpsd}) there is no power at the above mentioned frequencies. Therefore, it is probable that the 0.25 and 0.64\,$\mu$Hz peaks have been produced by instabilities in CoRoT but a stellar origin could still be possible.

\begin{figure}[!htbp]
\centering
\includegraphics[width=9.5cm]{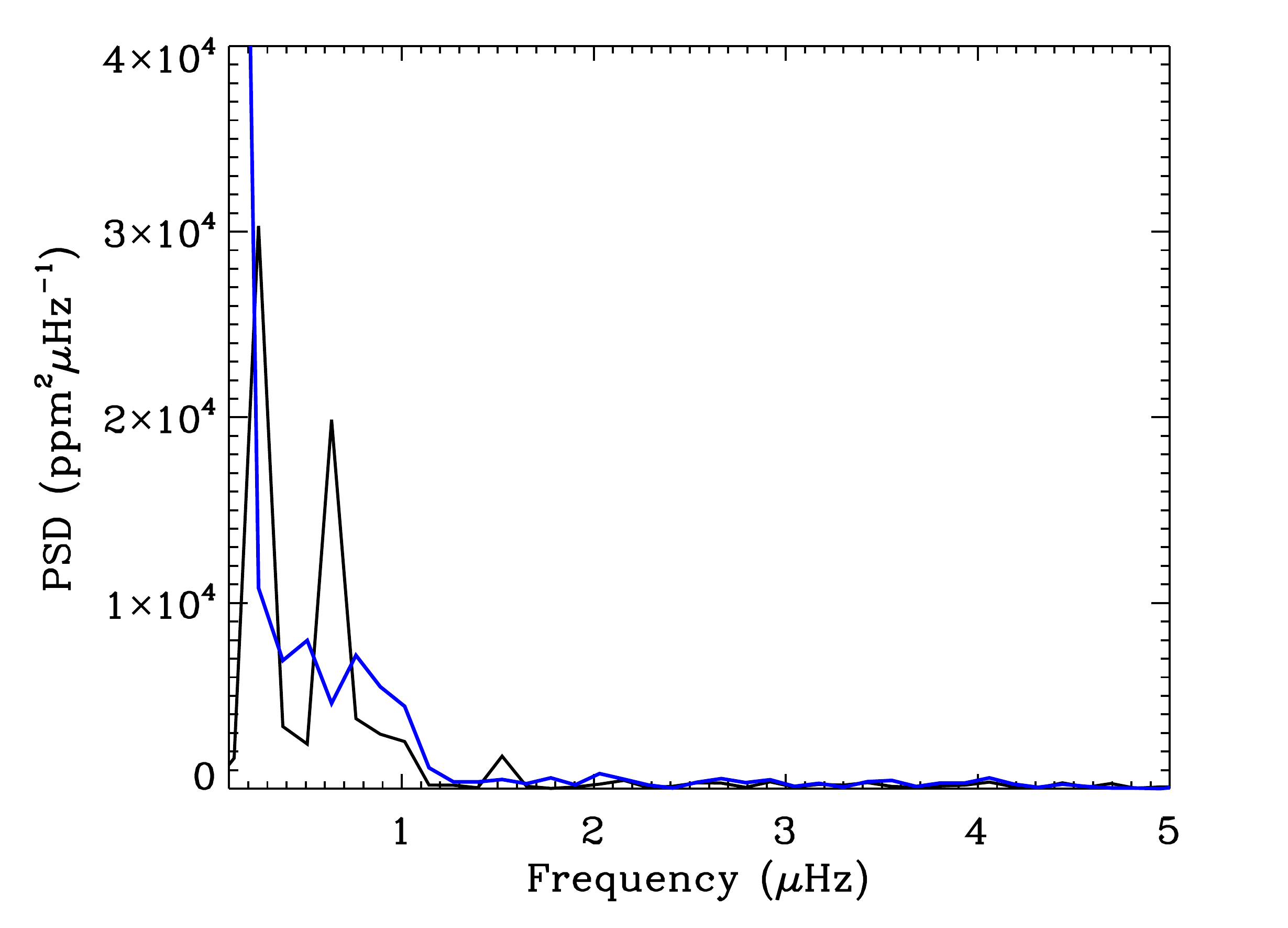}
\caption{Low-frequency part of the power spectrum between 0.1 and 5 $\mu$Hz. The black line correspond to the light curve showed in Fig~\ref{Fig:flux}. Two peaks are visible at 0.25 and 0.64 $\mu$Hz. The blue line correspond to the PSD of the light curve in which the jumps at 43.3 and 80.05 days have been corrected.}
\label{Fig:lowpsd}
\end{figure}

\section{Methodology followed to extract the mode parameters of HD~169392A}

Nine teams estimated the mode parameters for HD~169392A with two teams providing two sets of results from differing methods.  This gives a total of eleven sets of mode parameters.  We describe, albeit briefly, the methods used by the different teams to fit the power spectrum corrected from the background. A list of the methods used for each of the sets of mode parameters can be found in Table~\ref{fitters}.

\begin{table} 
\caption{Description of the fitting methods of each team. }             
\label{fitters}      
\centering                          
\begin{tabular}{c c c c }        
\hline
\hline                 
Fitter ID & Method & Splittings & Angle  \\    
\hline                        
   OB\tablefootmark{a} & MCMC global & Free & Free \\      
   JB\tablefootmark{b} & MAP global  & Free & Free \\ 
   GRD\tablefootmark{b} & MAP global & Free & Free \\ 
   RAG\tablefootmark{b} & MLE global & Free & Free \\ 
   RAG\tablefootmark{b} & MLE global & Fixed 0 & Fixed 0 \\ 
   SH\tablefootmark{c} & MLE pseudo-global  & Free & Free \\ 
   RH\tablefootmark{d} & MCMC global & Free & Free\\ 
   CR\tablefootmark{b} & MLE global & Free & Free \\ 
   CR\tablefootmark{b} & MLE global & Fixed 0 & Fixed 0 \\ 
   DS\tablefootmark{e} & MLE local & Free & Free \\ 
   GAV\tablefootmark{b} & MLE global & Free & Free \\ 
   
\hline                                   
\end{tabular}
\tablefoottext{a}{\citet{2008CoAst.157...98B}}\\
\tablefoottext{b}{\citet{AppGiz1998}}\\
\tablefoottext{c}{\citet{2011MNRAS.413..359F}}\\
\tablefoottext{d}{\citet{2011A&A...527A..56H}}\\
\tablefoottext{e}{\citet{2004A&A...413.1135S}}
\end{table}

All teams used a common philosophy for estimating mode parameters: maximise the likelihood function for a given model.  The models used by the teams are common in that all are composed from a sum of Lorentzian profiles to describe the modes of oscillation, plus a background to account for the instrumental and stellar noise (see Section~\ref{Sect:back}).  Each team used the frequency, height, linewidth, rotational splitting, and angle of inclination to describe a mode of oscillation \citep[e.g.][]{2008A&A...488..705A}.  All the teams fitted a single linewidth for each order. The main differences between teams arises from the method used to maximize the likelihood.  The three broad categories used are maximum likelihood estimation \citep[MLE,][]{AppGiz1998}, that has a variant called the maximum a posteriori method \citep[MAP,][]{2009A&A...506....7G}, and Markov Chain Monte Carlo \citep[MCMC,][]{2008CoAst.157...98B,2011A&A...527A..56H}. \\
The MLE or the MAP approaches were used by the majority of teams.  Briefly, MLE estimates the mode parameters by performing a multi-dimensional maximisation of the likelihood parameter. Typically formal one-sigma errors are determined from the Hessian matrix.  Like all Bayesian approaches, MAP techniques extend this concept by including {\it a priori} knowledge by implementing penalties on the likelihood parameter for defined regions of parameter space.  As for MLE, MAP errors are taken from the Hessian matrix.  Both MLE and MAP can be applied either globally, that is simultaneously fit to the entire frequency range \citep[e.g.][]{2008AN....329..485A}, or locally, considering parameters separately over one large separation \citep{2004A&A...413.1135S}. 
When coupled with MCMC, a Bayesian approach is a powerful way (but also time consuming) to extract the full statistical information with respect to the likelihood and our prior knowledge by mapping the probability density function (PDF) of each parameter. Contrary to MLE or MAP, MCMC techniques avoid to be trapped in local maxima and no assumptions on the nature of the probability density function are needed. Thus such approach provides more conservative and robust results, especially for the uncertainties as their estimation do not rely on the Hessian matrix, but on the cumulative distribution function, built thanks to the PDF. In the present paper, two fitters (OB, RH) used MCMC approaches with different priors. However, a careful choice of the priors was needed as it may have a strong impact on the best fitted parameters, especially at low signal-to-noise ratio. This star is a perfect example:  $\ell=0$ and $\ell=2$ frequencies lower than 870 $\mu$Hz (the three first radial orders) were barely fitted (posterior probability highly multimodal and non Gaussian) if no smoothness condition was applied onto the frequencies of these modes. A smoothness condition acts as a filter and avoid strong, erratic variations of frequencies from one order to another \citep{2012ApJ...745L..33B}.\\
Each of the three approaches can be applied in slightly different ways.  Two fitters (RAG \& CRR) produced two (MLE/MAP) results with and without the two parameters describing rotation rate and angle of inclination. In all cases for this analysis mode profile asymmetry was not included.  This is due to insufficient signal-to-noise ratio: asymmetry fitting requires very good signal-to-noise ratio and high resolution. 

The analysis has revealed the presence of mixed modes. As a consequence, to complement the fitting techniques, we have also used the asymptotic relation found by Goupil (private communication),
following the ideas originally developed by \citet{1989nos..book.....U} and tested by \citet{2012A&A...540A.143M}. This methodology is particularly suited to identify the presence of mixed modes in the power spectrum and to derive the measure of the gravity period spacing. This global seismic parameter provides a strong constraint on the core radiative region of the interior models.

Given eleven sets of fitters results, we required a statistically robust method of comparison to produce a final list of mode parameters for future modelling.  Here we adapted a previously used method \citep{2012A&A...543A..54A} to determine a confident list of mode frequencies.  This method used an iterative outlier rejection algorithm to populate two lists (a maximal and a minimal list) of modes which are then used to determine the ``best'' fitter who will provide the frequencies and is defined as the fitter with the smallest normalized rms deviation from the average frequency.\\
At each radial order, $n$, and degree, $l$, we compared $N$ different estimated frequencies.  As a first cut a visual inspection of all mode parameters and errors was performed.  Modes with clearly incorrect parameters (for example, very small/large linewidths or very large errors) were rejected.  Statistical outlier rejection was performed with Peirce's criterion \citep{1852AJ......2..161P,1855AJ......4...81G}, a method that is based on rigorous arguments.  Peirce's criterion was applied iteratively to each mode frequency set until no data points are rejected.   The method can be described with the following pseudo-code:
\begin{itemize}
\item COMPUTE mean $\bar{x}$ and rms $\sigma$ deviation from the sample $x$.
\item COMPUTE rejection factor $r$ from \citet{1855AJ......4...81G} assuming one doubtful observation.
\item REJECT: Reject data if $| x_{i} - \bar{x} \, | > r \sigma$. 
\item IF $n$ data are rejected THEN compute new $r$ assuming $n+1$ doubtful observations ELSE END.
\item GOTO REJECT.
\end{itemize}
With the results of the outlier rejection we populated the maximal and minimal list in the following way.  The minimal list was designed such that we had a very high level of confidence in the results included.  Previous works have used varying threshold criteria for inclusion in the minimal list.  \citet{2010ApJ...723.1583M} required that all fitters agree to place a frequency in the minimal list, i.e. $N$ results are accepted.  In an updated version of the recipe, \citep{2011A&A...534A...6C} and \citet{2011ApJ...733...95M} required that $N/2$ results are accepted.  Here we followed the $N/2$ acceptance criterion for the minimal list.  Modes forming the maximal list must have at least two results accepted by Peirce's criterion, because of this the maximal list must be treated with some caution.\\
In the maximal and minimal lists we selected the actual frequencies of the modes in each list.  We calculated the normalized root-mean-square (RMS) deviation of the $N$ teams with respect to the mean of the frequencies for each mode in the minimal list \citep[see Equation 2 of][]{2011ApJ...733...95M}.  The data set with the lowest normalized RMS deviation was declared the ``best'' fitter.  Two methods (MLE and MCMC) obtained very similar values of RMS and we decided to present in the following sections the results obtained by the MCMC, providing a reproducible set of frequency parameters.  

\begin{figure*}
  \centering
   \includegraphics[width=18cm]{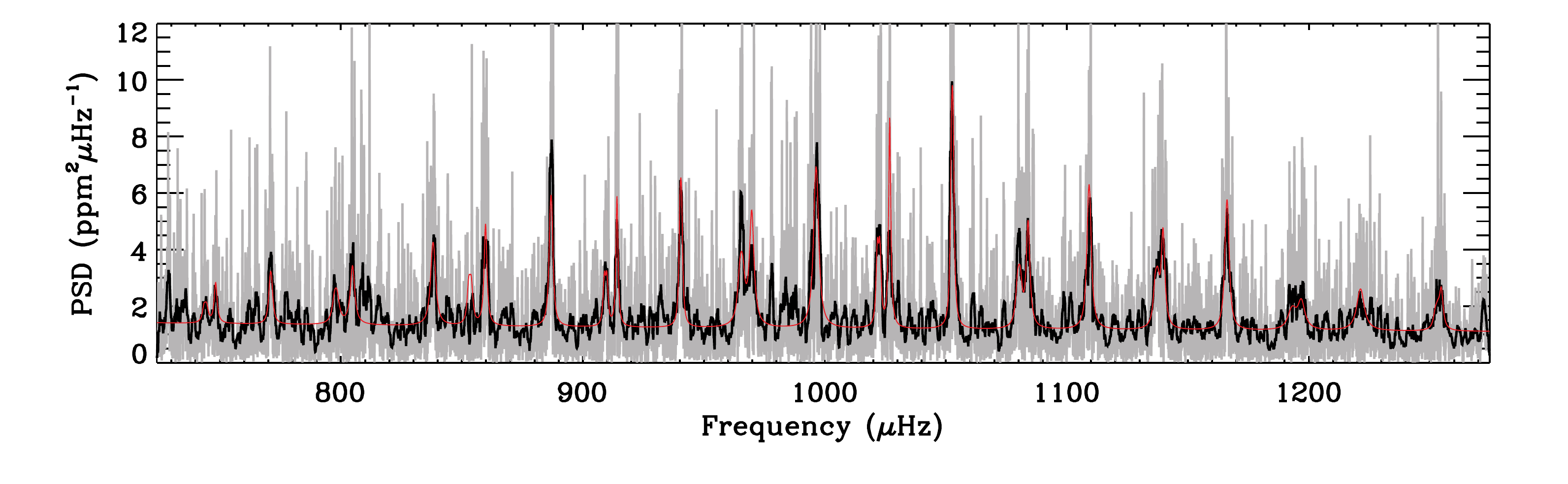}
	\includegraphics[width=18cm]{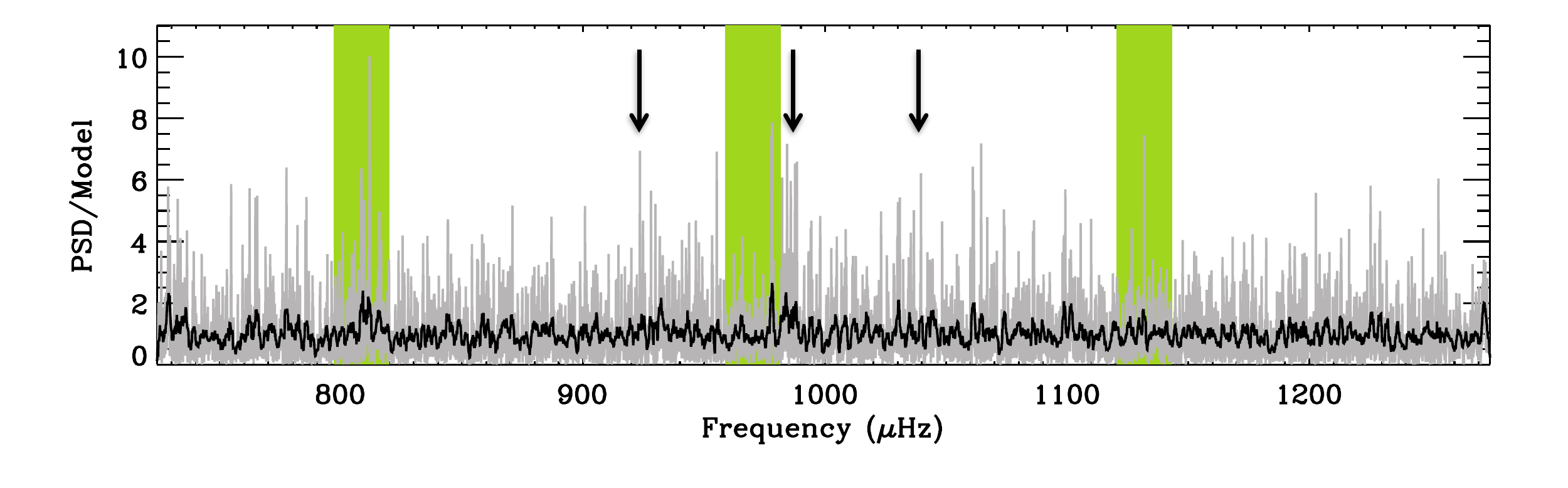}
      \caption{Upper panel: Power spectral density (PSD) of HD 169392 in the p-mode region at full resolution (grey) and smoothed over 15-bin wide boxcar (black).  The red line corresponds to the fitted spectrum from refit after the frequency comparison stage. Lower panel: Power spectral density for the p-mode region divided by the fitted model at full resolution (grey) and smoothed over 15 bins (black).  Regions highlighted in green cover the regions corresponding to the harmonics of the CoRoT orbital period such that $\nu = (n \times 161.7) \pm (m \times11.5) \; \rm \mu Hz$, where $n$ and $m$ are integers. The arrows show the positions where we would expect $\ell$\,=\,3 modes.
              }
         \label{fpsd}
\end{figure*}

\section{Results}

Following the methodology described in the previous section, we obtained the mode parameters presented in this section.

\subsection{Mode frequencies}

The comparison of the 11 sets of the modes frequencies provided by the 9 teams gave the minimal and maximal lists of frequencies. The fitter selected from the smallest RMS deviation (namely OB) used the MCMC algorithm and the model fitting the data is represented on Fig.~\ref{fpsd}. We obtained a minimal list of 25 modes. The maximal list contains 4 additional modes. 
 The lower panel of Fig.~\ref{fpsd} shows the ratio of the power spectrum to the fitted model.  The power spectrum can be thought as composed of a limit spectrum multiplied by a $\chi^{2}$ with a 2 degree-of-freedom noise function \citep[e.g.][]{1990ApJ...364..699A} and when smoothed over $k$ bins the noise function is distributed as a $\chi^2$ with $2k$ degrees of freedom \citep[e.g.][]{AppGiz1998}.  If the fitted model describes the limit spectrum well, then the ratio of the power spectrum to the model will be distributed as the noise function.  Hence, the power spectrum normalized by the model in the lower panel of Fig.~\ref{fpsd} reveals structures in the power spectrum not compensated by the fitted model in the region where the $\ell$ = 3 modes are expected (shown with arrows in the plot). 

\begin{table*}[hb]
\begin{center}
\caption{Minimal and maximal lists of frequencies for HD~169392 in $\mu$Hz obtained from 91 days of CoRoT observations with MCMC.\label{tbl-2}}
\begin{tabular}{ccccc}
\hline
\hline
  Order &$\ell$ = 0 & $\ell$ =1 & $\ell$ = 2  & $\ell$ = 3\\
\hline
    12 &  ... & ... & 743.22\,$\pm$\,1.34\tablefootmark{a} \\
     13  &   748.35\,$\pm$\,0.40\tablefootmark{a} &   770.96\,$\pm$\,0.37 &   797.97\,$\pm$\,0.76 &  ...\\
      14  &   804.50\,$\pm$\,0.35 &   838.21\,$\pm$\,0.38 &   853.50\,$\pm$\,0.85 &  ...\\
      15  &   859.70\,$\pm$\,0.38 &   886.83\,$\pm$\,0.23 &   909.34\,$\pm$\,0.46 &  ...\\
      16  &   914.14\,$\pm$\,0.27 &   940.52\,$\pm$\,0.27 &   965.45\,$\pm$\,0.41 &   930.52\,$\pm$\,1.87\tablefootmark{b}\\
      17  &   969.77\,$\pm$\,0.46 &   996.40\,$\pm$\,0.32 &  1022.18\,$\pm$\,0.27 &   986.08\,$\pm$\,1.20\tablefootmark{b}\\
      18  &  1026.79\,$\pm$\,0.27 &  1052.68\,$\pm$\,0.20 &  1080.04\,$\pm$\,0.44 &  1043.43\,$\pm$\,1.23\tablefootmark{b}\\
      19  &  1083.77\,$\pm$\,0.39 &  1109.19\,$\pm$\,0.36 &  1137.05\,$\pm$\,0.58 &  ...\\
      20  &  1139.76\,$\pm$\,0.42 &  1166.17\,$\pm$\,0.34 &  1193.29\,$\pm$\,0.97 &  ...\\
      21  &  1196.71\,$\pm$\,0.73 &  1221.76\,$\pm$\,1.38 &  1248.79\,$\pm$\,1.44\tablefootmark{a} &  ...\\
      22  &  1253.73\,$\pm$\,0.75\tablefootmark{a} &  ... &  ... &  ...\\

\hline
\end{tabular}

\tablefoottext{a}{Modes that belong to the maximal list.}\\
\tablefoottext{b}{Significant modes according to the MCMC method.}
\end{center}
\end{table*}

\begin{figure}[htb]
  \centering
   \includegraphics[width=9.5cm]{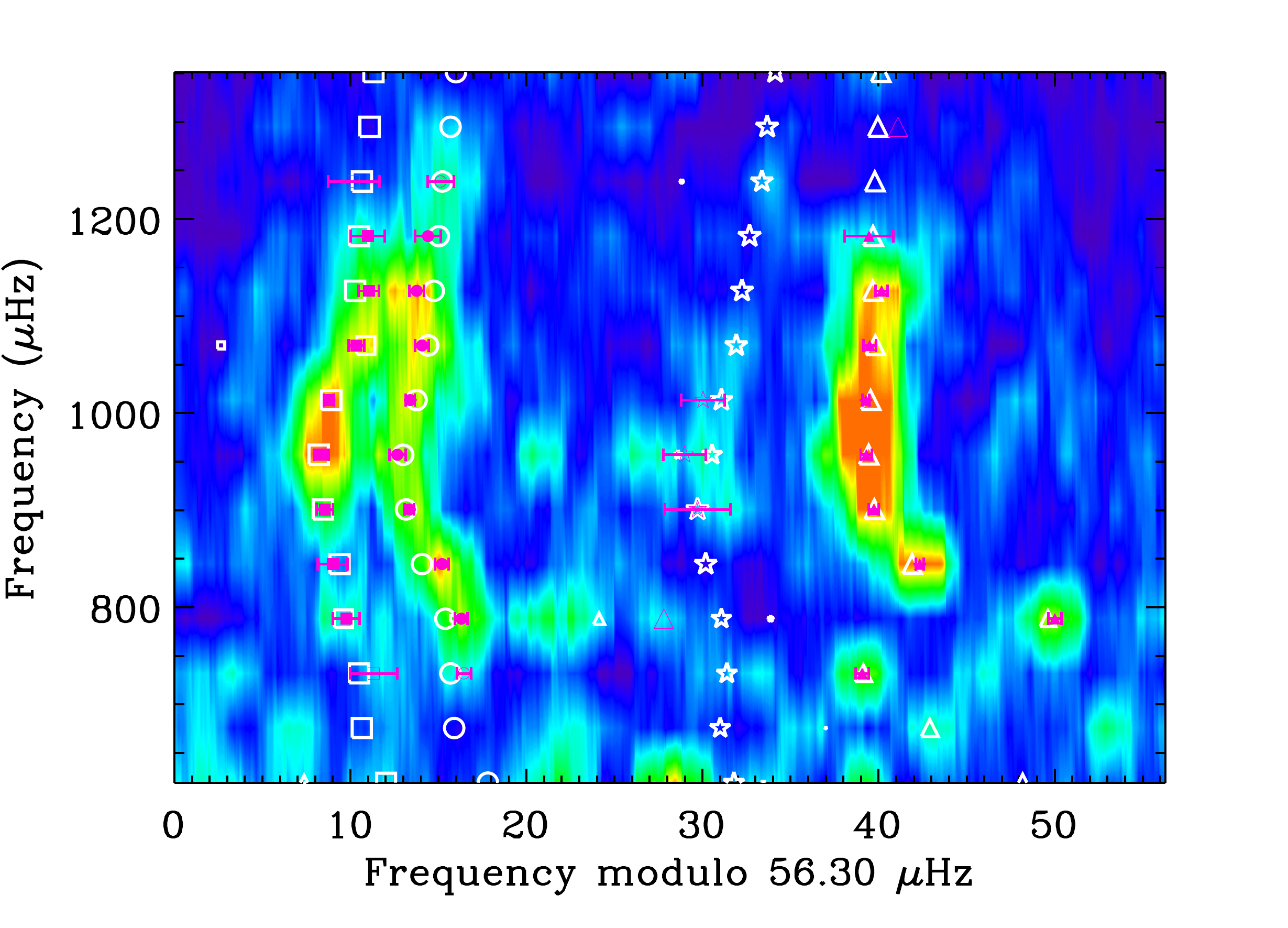}
      \caption{\'Echelle diagram of the CoRoT target, HD~169392A, with the minimal (close symbols) and maximal (open and filled pink symbols) lists of frequencies, and the frequencies from the best fit model of AMP (white symbols): $\ell$~=~0 (circles), $\ell$~=~1 (triangles), $\ell$~=~2 (squares), and $\ell$~=~3 (stars) as in Section~7.2. The pink triangles without uncertainties correspond to the mixed modes ($\ell$=1) mentioned in Section 7.3.
              }
         \label{minEd}
\end{figure}

\subsection{$\ell$=3 modes}

With the MCMC technique, we fitted the $\ell$=0, 1, 2, and 3 simultaneously. We checked that the frequencies of the modes $\ell$ =0, 1, and 2 remain unchanged. In Table~\ref{tbl-2}, we give the minimal and maximal lists of frequencies as well as the three $\ell$ = 3 modes that have a significant posterior probability. However, we remind the reader that these modes have to be taken very cautiously. By adding the fit of the $\ell$ = 3 modes, it slightly modified the final values of the amplitudes, widths, splittings, and inclination angle. The values listed in Table~\ref{tbl-4} correspond to the fitting of the four ridges.

\subsection{Mixed modes}
\label{Sect:mixed}

By looking at the \'echelle diagram (Fig.~\ref{minEd}), we notice an avoided crossing, corresponding to the presence of a mixed mode  \citep[e.g.][]{1975PASJ...27..237O,1977A&A....58...41A,2011A&A...535A..91D}.
The frequency pattern is affected by the coupling of the pressure wave in the outer convective region with a gravity wave in the radiative interior. Hence, we can use the formalism developed for mixed modes in giants \citep{2012A&A...540A.143M} to describe this pattern.

For p modes, the
second order correction of the asymptotic relation is related by a
curvature term $\alpha$:
\begin{equation}\label{tassoulp}
\nu_{\np,\ell} = \left(\np +{\ell \over 2}+\offsetp - d_{0\ell}
 + {\alpha \over 2}[ \np- \nmax ]^2 \right) \Dnu,
\end{equation}
where $\offsetp$ is the pressure phase offset, $d_{0\ell}$
accounts for the small separation, and $\nmax= \numax / \Dnu$. The
g modes are reduced to the first order term:
\begin{equation}
P_{\ng, \ell=1} = (|\ng| + \offsetg)\;  \Tg \ , \label{tassoulg}
\end{equation}
where $\ng$ is the gravity radial order, $\offsetg$ the gravity
phase offset and $\Tg$ the gravity period spacing of dipole modes.

\begin{table}
  \centering
  \caption{Mixed-mode parameters}\label{mmparam}
  \begin{tabular}{llr}
    \hline
    \hline
    \multicolumn{3}{c}{Asymptotic p modes}\\
    \hline
    Pressure phase offset         & $\varepsilon\ind{p}$ & 1.17\,$\pm$\,0.02 \\
    Small separation & $d_{01}$      &  0.03\,$\pm$\,0.02\\
    Curvature & $\alpha$  &  0.0029\,$\pm$\,0.0008\\
    \hline
    \multicolumn{3}{c}{Asymptotic g modes}\\
    \hline
    Gravity period spacing  & $\Delta\Pi_1$ &  476.9\,$\pm$\,4.3\, s\\
    Gravity phase offset    & $\varepsilon\ind{g}$ &  0.07\,$\pm\,0.03$\\
    \hline
    \multicolumn{3}{c}{Mixed modes}\\
    \hline
    Coupling & $q$   &  0.098\,$\pm$\,0.005\\
    \hline
  \end{tabular}
\end{table}

The coupling of the p and g waves is then expressed by a single
constant \citep{2012A&A...540A.143M}. Table \ref{mmparam} shows the
asymptotic parameters used for the fit. Compared to modes in
giants that have very large gravity orders (in absolute value),
mixed modes in a subgiant with $\Dnu\simeq 56\,\mu$Hz have small
$|\ng|$. Therefore, we cannot consider the asymptotic constant
$\offsetg$ to be 0.

The best fit of the radial ridge, obtained
from a global agreement over 9 orders, is provided with $\Dnu =
56.3\,\mu$Hz and $\offsetp=1.17$. We then infer a gravity period
spacing of the order of 476.9\,$\pm$\,4.3\,s, with a precision limited
by the uncertainty on the unknown gravity offset. Best fits are
obtained with $\offsetg=0.07\pm0.03$. In all cases, the coupling
is small: $q=0.098\pm0.005$. Two mixed modes are reported at
816 and 1336\,$\mu$Hz with gravity orders of $-2$ and $-1$,
respectively, corresponding to the period of {g modes} (the avoided crossing), $P\ind{g} = (|\ng|
+ 1/2 + \offsetg)\;  \Tg$ (Goupil et al. in prep.).
These mixed modes bring strong constraints on the age of a star as the coupling between the p- and g-mode cavities varies very fast with the stellar age, introducing strong constraints on the age of the star \citep[e.g.][]{2004SoPh..220..137C,2010ApJ...723.1583M,2012ApJ...749..152M}. The presence of the mixed modes in HD~169392A indicates that this star is quite evolved. 

Unfortunately, none of these two modes were fitted by the 9 teams. The peak at 816\,$\mu$Hz is surrounded by the orbital harmonics and without a proper modeling of these peaks, it is not possible to fit the mixed mode. The high frequency mixed mode at 1336\,$\mu$Hz has a too weak signal to be fitted.

\subsection{Frequency differences}

We fitted the frequencies of the $\ell$\,=\,0 modes against the order $n$ and according to Eq.~3, the slope is the mean large frequency separation. We obtained $\langle\Delta\nu\rangle$=55.98\,$\pm$\,0.05\,$\mu$Hz, which agrees with the value obtained from the global techniques (see Section~\ref{Sect:global}) and the mixed-mode asymptotic fit (Section~7.3). The variation of the mean large separation is shown in Fig.~\ref{large_sep}. We notice the dip for the modes $\ell$ = 1, due to the avoided crossing already mentioned. The small oscillation we observe is typical of the signature we can have for the helium's second ionization zone below the stellar surface. We would need more modes to be able to extract the position the second ionisation zone.


\begin{figure}
  \centering
   \includegraphics[width=9cm]{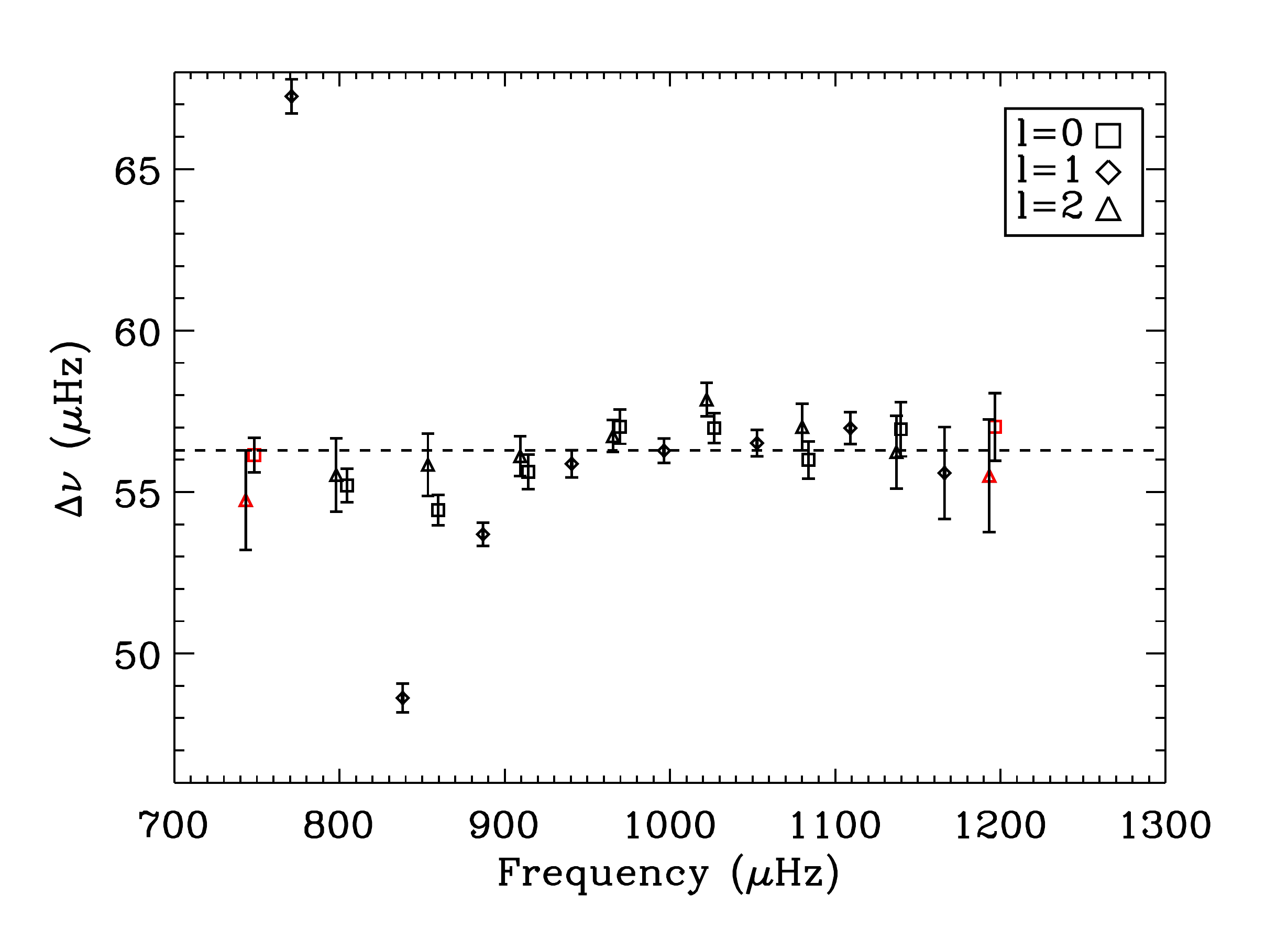}
      \caption{Large separation as a function of frequency for $\ell$ = 0, 1 and 2. The dashed line represents the mean large separation of the radial modes. The red squares at low and high frequency represent the values obtained using the modes of the maximal list.              }
         \label{large_sep}
\end{figure}

 We also computed the variation of the small separation between $\ell$ = 0 and 2 following \citet{2003A&A...411..215R} and it is shown in Figure~\ref{small_sep02}. The small separation presents a general decreasing trend (with a slope of $-$0.007), which is very similar to what we observe on the Sun. The mean value is around 4.14\,$\mu$Hz. This value fits perfectly the CD diagram presented by \citet{2011ApJ...742L...3W} where for subgiants, the small separation reaches approximately similar values for stars with very different masses. 


\begin{figure}
  \centering
   \includegraphics[width=9cm]{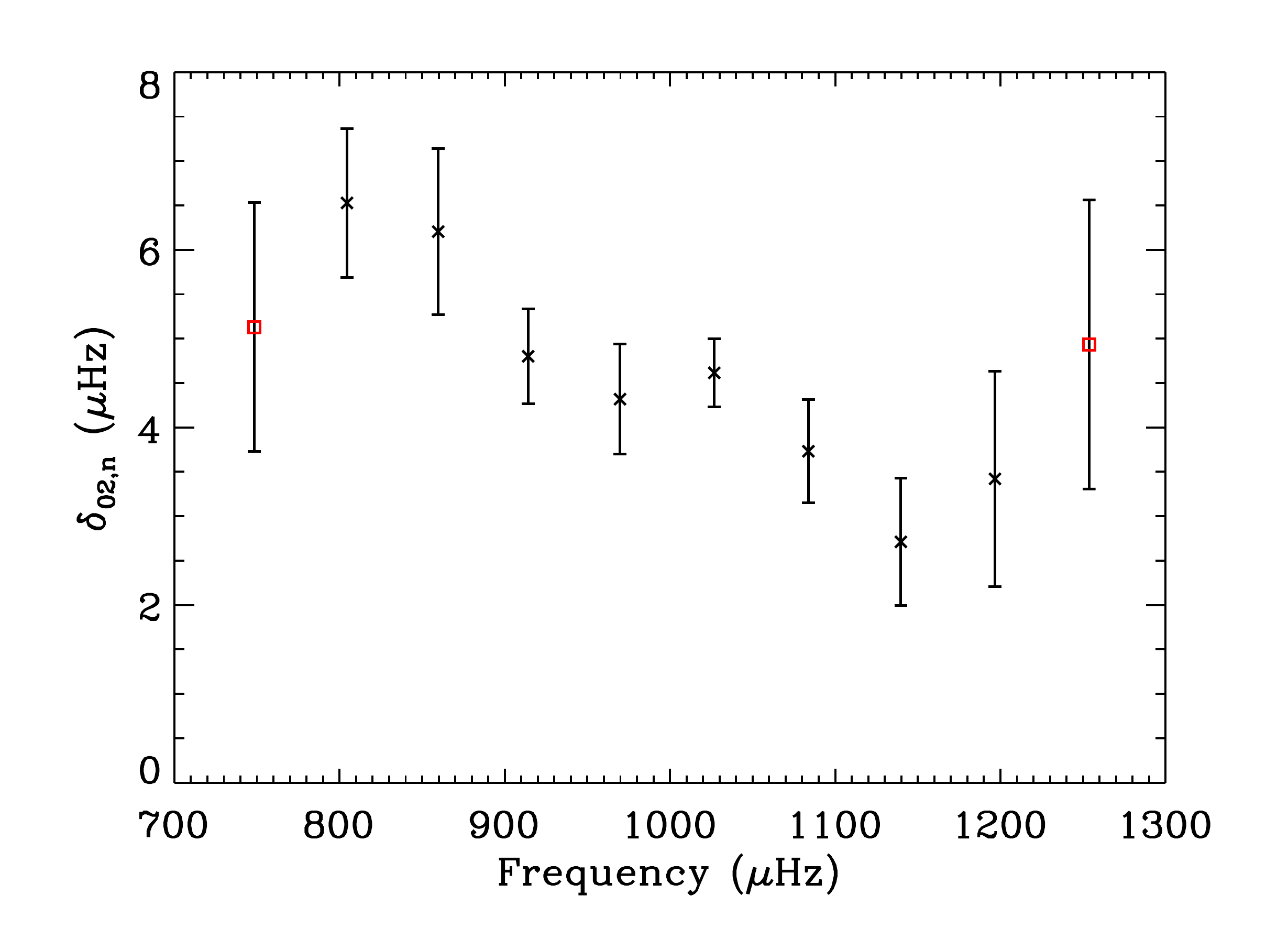}
      \caption{Small separation, $\delta_{02}$, as a function of frequency. The red squares at low and high frequency represent the values obtained using the modes of the maximal list. 
              }
         \label{small_sep02}
\end{figure}

\subsection{Linewidths}

The linewidths of the modes ($\Gamma_{\ell, n}$) are related to their lifetimes ($\tau_{\ell, n}$) following the formula for $\ell=0$ and order $n$:

\begin{equation}
\tau_{0,n} = (\pi \Gamma_{0,n})^{-1}
\end{equation}

The results from the MCMC method, which fitted a single linewidth for $\ell$=0, 1, and 2 modes per order, are given in Table~\ref{tbl-4} and represented in Fig.~\ref{lw}. 
We note that the trend is rather flat with a median value of 1.8\,$\mu$Hz, with a little increase with frequency. The first and last points were obtained with the frequency of the maximal list of modes. This behavior is similar to the one found in the Sun in the plateau region and the higher frequencies. We also observe a small oscillation followed by a small increase of the linewidth, which is also seen in other G-type stars observed by CoRoT \citep[e.g.][]{2010A&A...515A..87D,2011A&A...530A..97B} .

Several works have been done to show the dependence of the linewidths with the effective temperature of the star as $\Gamma \sim T_{\rm eff}^s$. \citet{2009A&A...500L..21C} derived a relation where the mean linewidth scales as $T_{\rm eff}^{-4}$ for temperatures ranging between 6800~K and 5300~K. \citet{2011A&A...529A..84B} re-evaluated the exponent to be 16\,$\pm$\,2 using 5 main-sequence stars observed by CoRoT and the Sun. The latest work by \citet{2012A&A...537A.134A} based on the analysis of 42 cool main-sequence stars and subgiants observed by {\it Kepler} refined the expression to $\Gamma \sim T_{\rm eff}^{15.5}$. 
For HD169392, we measure the linewidth at maximum height as 1.41\,$\mu$Hz ($\ell=0$, $n=18$) that agrees very well with Fig.~2 of \citet{2012A&A...537A.134A} using $T_{\rm eff}$\,=\,5885\,K and with the theoretical computation by \citet{2012A&A...540L...7B}.  Finally, \citet{2012arXiv1205.4023C} analysed the linewidth of red-giant stars in clusters and found that by taking into account main-sequence and subgiant stars, $\Gamma$ varies exponentially with $T_{\rm eff})$, which gives 1.66\,$\mu$Hz, still in agreement with the observed value.

\begin{figure}
  \centering
   \includegraphics[width=9cm]{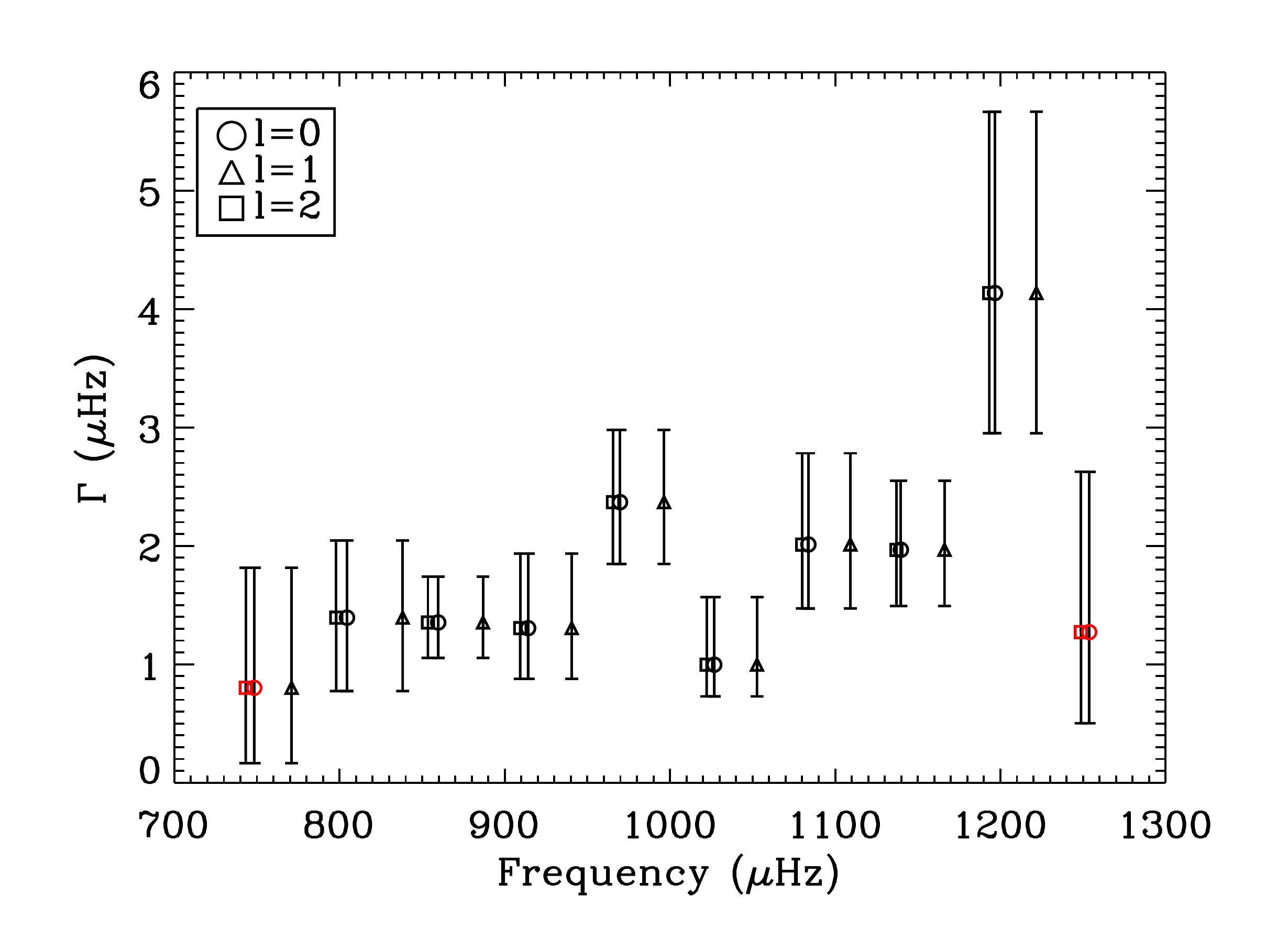}
      \caption{Linewidths of $\ell$ = 0 modes as a function of frequency. Same legend as in Fig.~\ref{small_sep02}. A single linewidth has been fitted for each order. }
         \label{lw}
\end{figure}

\begin{table*}
\begin{center}
\caption{Amplitudes and linewidths for all the modes of Table~\ref{tbl-2} computed with the MCMC. }
\begin{tabular}{cccccccc}
\hline
\hline
$l$ & $n$ & Linewidth ($\mu$Hz) & $+\sigma$ & $-\sigma$ & Amplitude (ppm)& $+\sigma$ & $-\sigma$ \\
\hline

0  &  13  &     0.80 & 1.02 & 0.64  &     1.63 & 0.33 & 0.32\\
0  &  14  &     1.39 & 0.65 & 0.62  &     2.40 & 0.30 & 0.32\\
0  &  15  &     1.35 & 0.39 & 0.30  &     2.83 & 0.29 & 0.27\\
0  &  16  &     1.30 & 0.63 & 0.43  &     2.81 & 0.31 & 0.26\\
0  &  17  &     2.37 & 0.61 & 0.52  &     3.82 & 0.30 & 0.29\\
0  &  18  &     1.00 & 0.57 & 0.27  &     3.72 & 0.36 & 0.37\\
0  &  19  &     2.01 & 0.77 & 0.54  &     3.39 & 0.31 & 0.29\\
0  &  20  &     1.97 & 0.58 & 0.47  &     3.23 & 0.24 & 0.25\\
0  &  21  &     4.14 & 1.53 & 1.18  &     2.54 & 0.25 & 0.27\\
0  &  22  &     1.27 & 1.35 & 0.77  &     1.36 & 0.29 & 0.36\\
1  &  13  &     0.80 & 1.02 & 0.64  &     1.94 & 0.38 & 0.40\\
1  &  14  &     1.39 & 0.65 & 0.62  &     2.84 & 0.37 & 0.37\\
1  &  15  &     1.35 & 0.39 & 0.30  &     3.37 & 0.32 & 0.31\\
1  &  16  &     1.30 & 0.63 & 0.43  &     3.35 & 0.34 & 0.32\\
1  &  17  &     2.37 & 0.61 & 0.52  &     4.53 & 0.34 & 0.33\\
1  &  18  &     1.00 & 0.57 & 0.27  &     4.40 & 0.42 & 0.38\\
1  &  19  &     2.01 & 0.77 & 0.54  &     4.04 & 0.33 & 0.33\\
1  &  20  &     1.97 & 0.58 & 0.47  &     3.83 & 0.31 & 0.31\\
1  &  21  &     4.14 & 1.53 & 1.18  &     3.02 & 0.31 & 0.33\\
2  &  12  &     0.80 & 1.02 & 0.64  &     1.34 & 0.28 & 0.27\\
2  &  13  &     1.39 & 0.65 & 0.62  &     1.96 & 0.28 & 0.27\\
2  &  14  &     1.35 & 0.39 & 0.30  &     2.33 & 0.25 & 0.24\\
2  &  15  &     1.30 & 0.63 & 0.43  &     2.32 & 0.25 & 0.23\\
2  &  16  &     2.37 & 0.61 & 0.52  &     3.13 & 0.25 & 0.25\\
2  &  17  &     1.00 & 0.57 & 0.27  &     3.04 & 0.32 & 0.29\\
2  &  18  &     2.01 & 0.77 & 0.54  &     2.79 & 0.26 & 0.26\\
2  &  19  &     1.97 & 0.58 & 0.47  &     2.64 & 0.27 & 0.24\\
2  &  20  &     4.14 & 1.53 & 1.18  &     2.08 & 0.23 & 0.25\\
2  &  22  &     1.69 & 1.42 & 0.91  &     0.99 & 0.27 & 0.29\\
3  &  16  &     1.30 & 0.63 & 0.43  &     0.65 & 0.07 & 0.06\\
3  &  17  &     2.37 & 0.61 & 0.52  &     0.89 & 0.07 & 0.07\\
3  &  18  &     1.00 & 0.57 & 0.27  &     0.86 & 0.08 & 0.09\\
\hline
\label{tbl-4}
\end{tabular}
\end{center}
\end{table*}

\subsection{Amplitudes}

From the heights and linewidths of the modes, we can compute the rms amplitude according to the following expression:

\begin{equation}
A_{\ell, n}=\sqrt{\frac{\pi \Gamma_{\ell, n} H_{\ell, n}}{2}}
\end{equation}

\noindent where $\Gamma_{\ell, n}$ is the width of the mode ($\ell$, $n$) and $H_{\ell, n}$ is the height of that mode in a single-sided power spectrum (hence the factor 2 in the numerator).
The values obtained are listed in Table~\ref{tbl-4} and shown in Fig.~\ref{Amplitude}. We can see the profile of the p-mode bump. The maximum amplitude of the modes is of 3.82\,$\pm$0.30\,ppm, which agrees with the value obtained in Section 4.1 (3.61\,$\pm$\,0.35\,ppm). Due to the imperfect observational window (with a duty cycle of $\sim$\,0.83), there is a non-negligible leakage of mode power in small aliases. We need to correct the amplitude for this effect and this leads to an observed value of 4.12\,$\pm$\,0.32\,ppm. To take into account the response function of CoRoT, we convert the amplitude to the bolometric one by computing $R_{\rm osc}$=7.87 and $R_{\ell=0}$=4.44 \citep[see][for details]{2009A&A...495..979M}, which gives the bolometric correction $c_{\rm bol}$=0.90\,$\pm$\,0.01. Finally, since HD~169392 is a binary, a part of the observed total flux comes from the secondary component. With the magnitude quoted in Section 1, we find $L_A /L_B$ =3.9, which allows us to conclude that $\sim$\,20\% of the flux is dimmed by the second component. Therefore, the final value for the maximal amplitude of the modes is $A_{\rm max}^{\rm obs}$\,=\,4.45\,$\pm$\,0.34\,ppm.

Several expressions have been developed the last few years to compute the maximum amplitude of the modes. According to \citet{2007A&A...463..297S}, the theoretical value of the maximum amplitude is obtained with:

\begin{equation}
A_{\rm max}= \Big( \frac{L/L_{\odot}}{M/M_\odot} \Big)^s \sqrt{\frac{5777}{T_{\rm eff}}} A_{\odot, {\rm max}}
\end{equation}

 \noindent with $s$=0.7 and according to \citet{2009A&A...495..979M}, $A_{\odot, \rm max}$=2.53\,$\pm$\,0.11\,ppm (in rms). Using the values of $R$ and $M$ from the AMP model (see Section 8), we obtain 6.02\,$\pm$\,0.60\,ppm.
 
 We computed the theoretical value with the updated scaling relation derived by \citet{2011A&A...529L...8K}, which depends on the lifetime of the modes, $\tau_{\rm osc}$:
 
  \begin{equation}
A_{\rm max} \propto \frac{L \tau_{\rm osc}^{0.5}}{(M/M_\odot)^{1.5} T_{\rm eff}^{2.25+r}}
\end{equation}

 We took $r$=1.5 as adopted by \citet{2008Sci...322..558M} and we obtain 4.32\,$\pm$\,1.10\,ppm, which is much closer to the observed amplitude.
 
\citet{2011ApJ...743..143H} used several hundreds of stars (red giants and main-sequence stars) to derive an empirical relation, slightly different from the previous one and that is similar to the results obtained from red giants observed in clusters \citep{2011ApJ...737L..10S}. This relation is more dependent on the mass of the star:
 
 \begin{equation}
A_{\rm max}= \frac{(L/L_{\odot})^s}{(M/M_\odot)^t} \frac{5777}{T_{\rm eff}} A_{\odot, {\rm max}}
\end{equation}

\noindent with $s$=0.838, $t$=1.32, and $A_{\odot, \rm max}$=2.53\,$\pm$\,0.11\,ppm (in rms), leading to 6.58\,$\pm$\,0.79\,ppm for \citet{2011ApJ...743..143H}. \citet{2011ApJ...737L..10S} found slightly different parameters: $s$=0.95 and $t$=1.8 (in the adiabatic case), which gives $A_{\rm max}$\,=\,7.21\,$\pm$\,0.92\,ppm.

The observed value is at the lower limit of the theoretical ones within  3-$\sigma$ for modelings in equations (7) and (9) and agrees within 1-$\sigma$ for the prediction from equation (8).


\begin{figure}
  \centering
   \includegraphics[width=9cm]{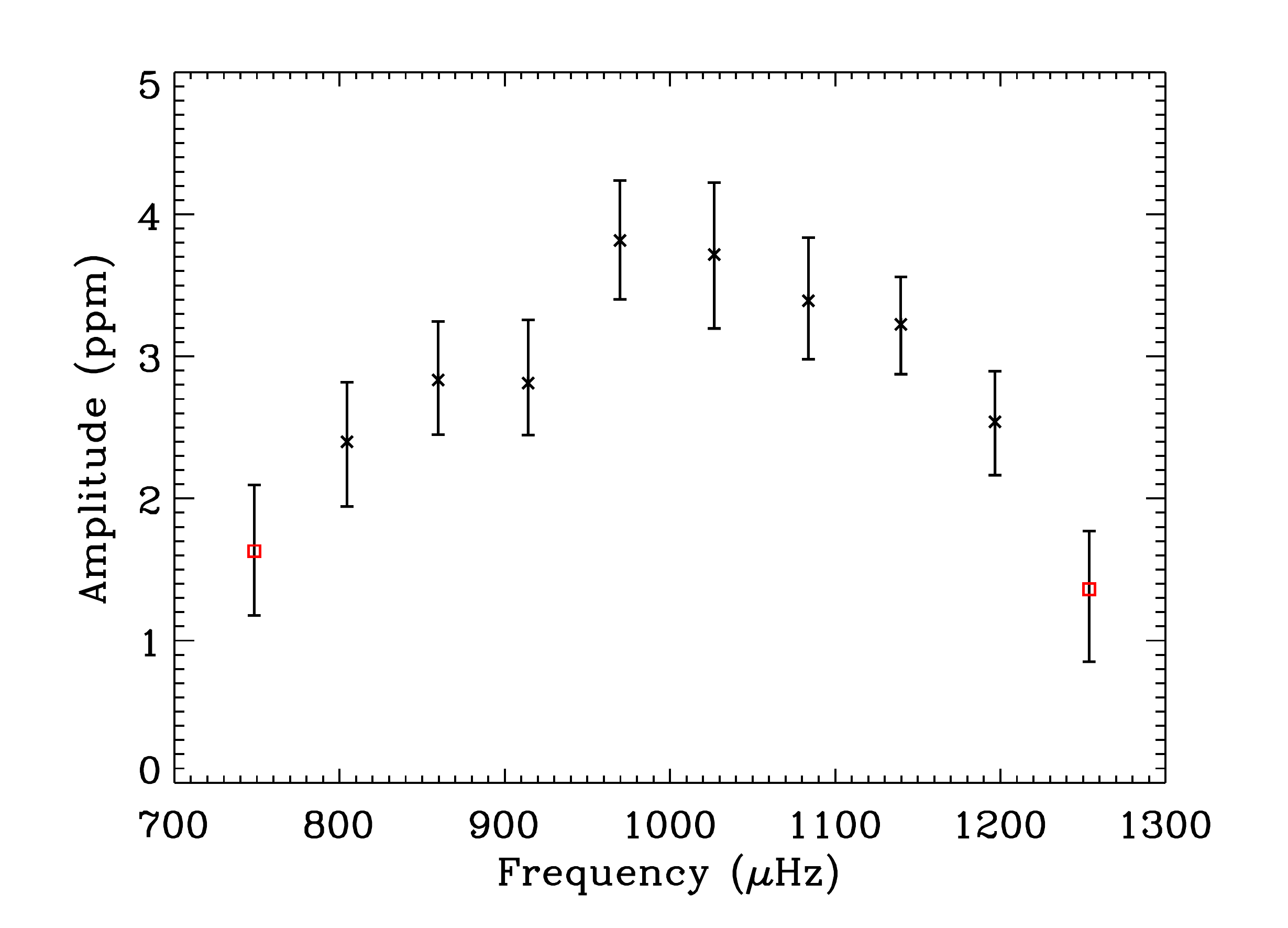}
      \caption{Amplitude of $\ell$ = 0 modes as a function of frequency. Same legend as in Fig.~\ref{small_sep02}.
              }
         \label{Amplitude}
\end{figure}

\subsection{Relative visibilities}


Accurate estimations of the relative visibilities of non-radial modes are important for our knowledge of the physics of the stellar atmospheres as they depend on the stellar limb darkening and the observed wavelengths. Although well defined in the Sun at different heights in the solar photosphere \citep{2011A&A...528A..25S}, their measurements are more critical for other stars due to lower signal-to-noise ratio and shorter time series. Nevertheless, the mode visibilities were successfully measured in two of the CoRoT stars: HD~52265 \citep{2011A&A...530A..97B} and HD~49385 \citep{2010A&A...515A..87D}. The uncovered visibilities quantitatively agree well with the theoretically calculations of limb-darkening functions from \citet{2011A&A...531A.124B}, who also showed a dependence on $T_{\rm eff}$. 

Estimations of the relative visibilities, $V_{l}^2$, of $\ell=1$, 2, and 3 modes were obtained for HD169392A. Consistent values within the error bars were obtained among the different peak-fitting techniques. Some tests using different inclination angles and rotational splittings showed that the uncovered visibilities give similar results within the uncertainties. The relative visibility of the modes $\ell=1$, 2, and 3 measured by the MCMC analysis are given in Table~\ref{tbl-5}, as well as the median values of their distributions and their associated 1-$\sigma$ values.

\subsection{Splitting and inclination angle}\label{sect:split}

We took the step to check that mode parameters were independent of the assumed angle of inclination.  Multiple fits with a fixed rotational splitting  range of fixed angles were tested based on the MCMC results. 
 The mode frequencies returned by the various fits were virtually independent of the angle used.  A last check was  done by directly looking at the correlation map between width, inclination angle and splittings from the MCMC technique. The mode linewidths and amplitudes returned by these fits showed a small dependence on the angle of inclination and the splittings, but at a much lower level than the 1-$\sigma$ error bars over the range $15\,\degr \leq i \leq 85\,\degr$ -- the range limits suggested by MCMC results.\\

Fig.~\ref{fig:cor_map} shows the correlation map of the splitting and the inclination (two-dimensional probability density function).
We notice two maxima of almost equal heights into the joint-posterior probability density function.  This bi-modality suggests that two different values for the joint-parameters are possible. However, the two solutions are compatible within 2\,$\sigma$ and therefore are not statistically different (cf. Table~\ref{tbl-5}). In order to have a more stringent determination, one would need additional constraints, such as the signature of the surface rotation, unfortunately unavailable with the present dataset (see Section~\ref{rotation}).

 The product of the two quantities, which gives the projected splitting, $\nu_{s} \sin i$, is well defined as seen by the curved ridge and the uncertainties presented in Table~\ref{tbl-5}.  However, as anticipated by \citet{BalGar2006}, the robust estimation of the individual quantities of angle and splitting is hampered by two competing solutions, the first with relatively low angle and high splitting, and the second with high angle and lower splitting.  Our inability to discern between the two solutions prevents us from reporting the desired 'robust' result.

\begin{figure}
  \centering
   \includegraphics[width=9cm, trim=4cm 3cm 0.5cm 2cm]{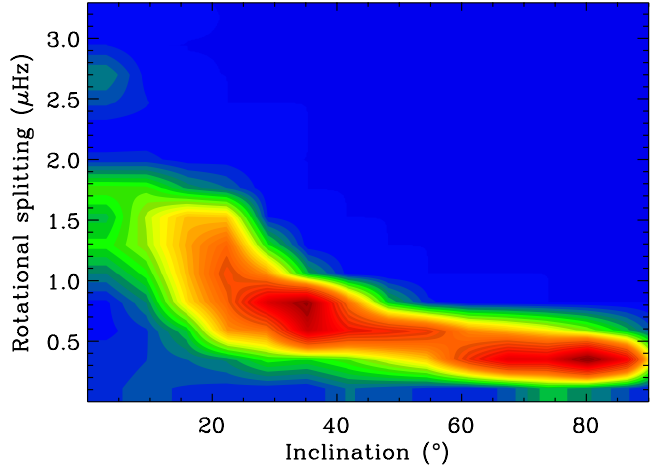}
      \caption{Correlation map between rotational splitting and inclination. The red (resp. yellow) region correspond to the 1-$\sigma$ (resp. 2-$\sigma$) confidence area.
              }
         \label{fig:cor_map}
\end{figure}

\begin{table}

\caption{Medians and standard deviations for the inclination angle ($i$), the rotational splitting ($\nu_s$), the projected splitting ($\nu_s \sin i$) and the relative visibilities of the $\ell=1$, 2, and 3 modes ($V_{l=1}$, $V_{l=2}$,  $V_{l=3}$) obtained from MCMC. These statistical information are deduced from the joint-probability density function of the inclination/splitting and from the probability density function for the other parameters.}

\begin{tabular}{ccccccc}
\hline
\hline
	 					      & $i$ ($^\circ$) & $\nu_s$ ($\mu$Hz) & $\nu_s \sin i$ ($\mu$Hz) & $V^2_{l=1}$ & $V^2_{l=2}$ & $V^2_{l=3}$\\ \hline 
	Median           & $36$						  &   $0.62$						&     $0.38$ 				 	&  $1.41$      &  $0.68$		 & $0.15$ \\
	$-\sigma$        & $20$							&   $0.23$						&			$0.16$ 					&  $0.17$ 		 & 	$0.11$     & $0.07$ \\
	$+\sigma$        & $36$							& 	$0.59$						&			$0.11$ 					&  $0.20$ 		 &  $0.11$     & $0.08$ \\ 
\hline
\end{tabular}
\label{tbl-5}
\end{table}

\section{Modeling HD~169392A}

With the spectroscopic and asteroseismic information retrieved in the previous sections, we modeled HD~169392A with the Asteroseismic Modeling Portal \citep[AMP,][]{2009ApJ...699..373M}. AMP is a genetic algorithm that aims at optimizing the match between the observed and the modeled frequencies of the star in addition to the spectroscopic constraints. The models are computed by the Aarhus STellar Evolution Code \citep{2008Ap&SS.316...13C}, which is also used to build the standard model of the Sun. It includes the OPAL 2005 equation of state \citep{2002ApJ...576.1064R} and the most recent OPAL opacities \citep{1996ApJ...464..943I} associated with the low-temperature opacity table of \citet{1994ApJ...437..879A}. Finally, the convection is treated according to the mixing-length theory \citep{1958ZA.....46..108B}, and diffusion and gravitational settling of helium follow the prescription by \citet{1993ASPC...40..246M}. A surface correction was applied to the frequencies following the prescription by \citet{2008ApJ...683L.175K}. AMP has already modeled a large number of stars \citep[e.g.][]{2010ApJ...723.1583M,2012ApJ...749..152M,2012ApJ...748L..10M} for a given physics.

For the observables, we used the spectroscopic constraints from Table~\ref{tab:param} and the mode frequencies from the minimal list (Table~\ref{tbl-2}). The best fit model obtained has a $\chi^2$ of 1.02. Since seismic and spectroscopic observables are used, we also computed two separate normalized $\chi^2$: $\chi^2_{\rm spec}$=0.22 and $\chi^2_{seis}$=1.43. 

The $\chi^2_{\rm spec}$ shows a good agreement between the observables and the best fit model. In Figure~\ref{minEd}, the modeled frequencies are superimposed on the \'echelle diagram obtained with CoRoT, and we can see that the best fit model frequencies match well the observations. We can also notice that the maximal list frequencies, the $\ell$\,=\,3 modes included, are well reproduced by the model, suggesting that these frequencies probably correspond to real signal. 

We obtained a mass of M\,=\,1.15\,$\pm$\,0.01\,M$_{\odot}$, a radius of R\,=\,1.88\,$\pm$\,0.02\,R$_{\odot}$, and an age $\tau$\,=\,4.33\,$\pm$\,0.12\,Gyr, where the uncertainties quoted are internal errors. 

Using the parallax $\pi = 13.82\,\pm\,1.2\,mas$, we obtained a luminosity $L\,=\,4.16\,\pm\,0.6\,L_{\odot}$ while the value from the best-fit model of AMP is 4.18\,$\pm$\,0.01\,$L_{\odot}$, which is a very good agreement.

\begin{table*}
\center
\caption{Modeling results from AMP, direct method + IRFM, and grid-based method +IRFM. The uncertainties for the AMP best fit model are the internal errors.}
\begin{tabular}{ccccccc}
\hline
\hline
Method & M (M$_\odot$) & R (R$_\odot$) & $\tau$ (Gyr) & \logg (dex) & $T_{\rm eff}$ (K) & $L/L_{\odot}$\\ \hline 
AMP & 1.15\,$\pm$\,0.01 & 1.88\,$\pm$\,0.02 & 4.33\,$\pm$\,0.12 & 3.95\,$\pm$\,0.01  & 6016\,$\pm$\,5 & 4.18\,$\pm$\,0.01  \\
Direct method + IRFM & 1.30\,$\pm$\,0.23 & 1.93\,$\pm$\,0.13 & - & 3.97\,$\pm$\,0.02 & 6002\,$\pm$\,77 & 4.45\,$\pm$\,0.67\\
Grid-based method + IRFM & 1.25\,$^{+0.05}_{-0.03}$\, & 1.96\,$\pm$\,0.02 & - & 3.95\,$\pm$\,0.01 & 6002\,$\pm$\,77 & 4.51\,$\pm$\,1.04\\
\hline
\end{tabular}
\label{tbl-6}
\end{table*}

We have also followed the methodology developed by \citet{silva2012} to couple the Casagrande et al. 2010 implementation of the InfraRed Flux Method (IRFM) with asteroseismic analysis. These technique provides results using both the direct and grid-based methods, as well as a determination of the distance to the target. The results are in complete agreement with the AMP ones for the \logg and $T_{\rm eff}$ (see Table~\ref{tbl-6}). The mass and radius agree with the first method within 1\,$\sigma$ and with the second one within 2\,$\sigma$. The distance determined by these techniques is d=70\,$\pm$\,5.3 pc, in excellent agreement with parallax measurements.

The best fit model of AMP confirms that  this is indeed a subgiant that has exhausted its central hydrogen and has no convective core. As seen on the HR diagram with the evolutionary track for this model (Figure~\ref{hrdiag}), it has not yet started ascending the red giant branch.

\begin{figure}
  \centering
   \includegraphics[width=9cm, trim=2cm 1cm 0cm 1cm]{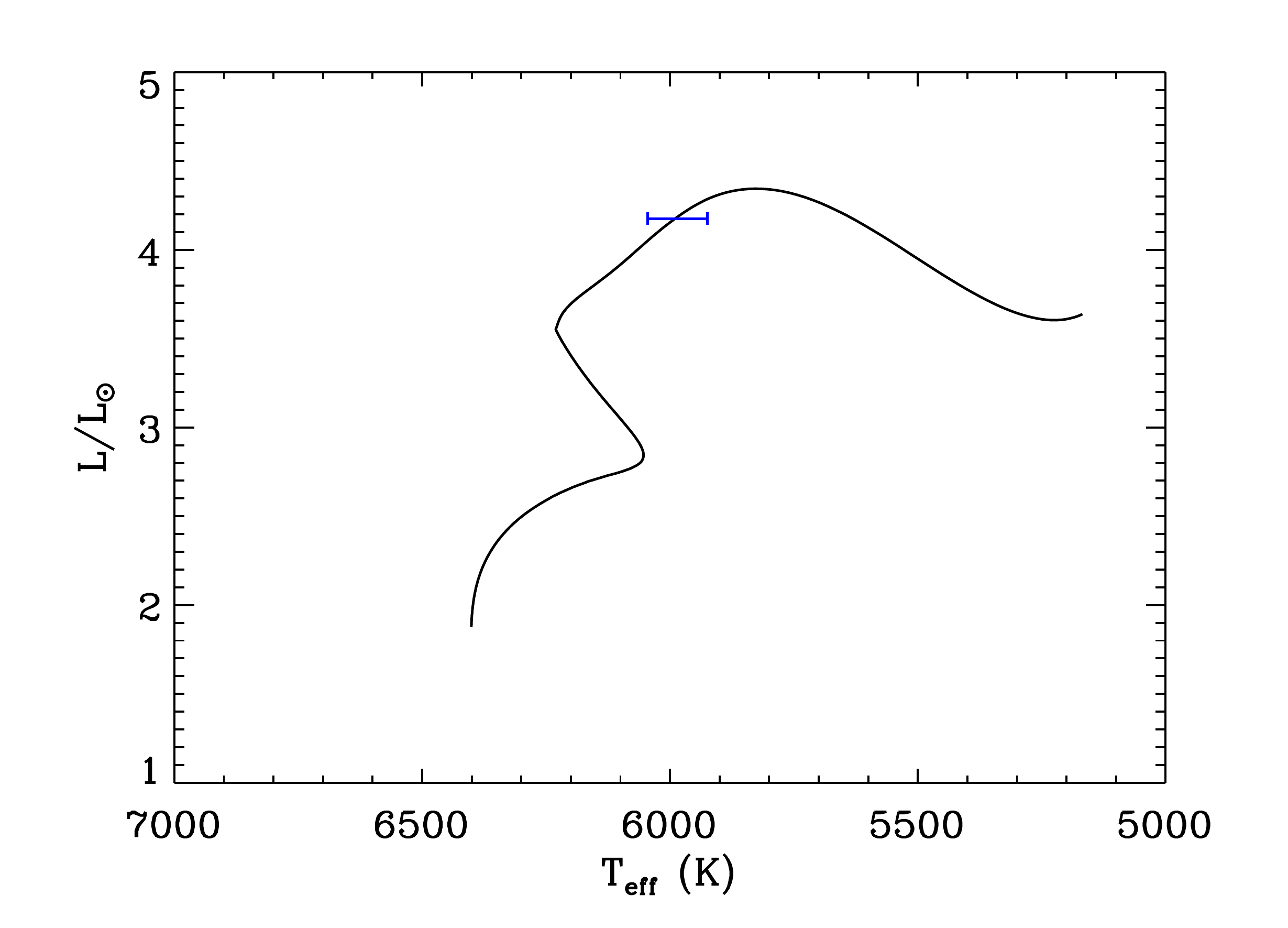}
      \caption{Evolutionary track computed by AMP and position of the best fit model with the spectroscopic uncertainty (blue symbol).}
              
         \label{hrdiag}
\end{figure}

\section{Conclusion}

In the present analysis, we have derived the fundamental parameters of the two components of the binary system using HARPS observations. We obtained very close values of metallicity for the two components. This is quite similar to what has been observed for other binary systems \citep[e.g. 16Cyg][]{2012ApJ...748L..10M}, suggesting that the A and B components were formed from the same cloud of gas and dust. None of the two components have significant amount of Li.

The analysis of 91 days of the CoRoT observations did not allow us to detect a significant seismic signature of the B component. This result agrees with the low detectability probability of 2\% for the B component.

We measured the global parameters of the modes of HD~169392A leading to $\dnumoy$ of 56.98\,$\pm$\,0.05\,$\mu$Hz and $\nu_{\rm max}$ of 1030\,$\pm$\,55\,$\mu$Hz making this star very similar to another CoRoT target, HD~49385 \citep{2010A&A...515A..87D}.
 
 Nine groups reported p-mode parameters and the comparison of the results led to a minimal and maximal lists of frequencies. Some significant power was found in the region where we would expect $\ell$\,=\,3 modes. We fitted the $\ell$\,=\,0, 1, 2, and 3 modes all together with the MCMC technique and obtained that three modes $\ell$\,=\,3 are significant according to the posterior probability. We analysed the mixed modes of the star using their asymptotic properties and detected two avoided crossings at 816\,$\mu$Hz and 1336\,$\mu$Hz. From these avoided crossings, we derive the gravity spacing $\Delta\Pi_1$\,=\,477\,$\pm$\,5\,s. We conclude that this star is quite evolved but not yet on the red-giant branch.
 
 We derived the amplitude of the modes, which is smaller than the predicted amplitudes within 1- to 3-$\sigma$ depending on the theoretical formula applied. The linewidths of the modes obtained are coherent with the dependence on the effective temperature derived by \citet{2012A&A...537A.134A} and \citet{2012arXiv1205.4023C}.
 
 The study of the splittings and inclination angle led to two possible solutions. Either we obtain an inclination angle of $20-40\,\degr$ with splittings of $0.4-1.0$\,$\mu$Hz, or we have an inclination angle of $55-85\,\degr$ and splittings of $0.2-0.5$ $\mu$Hz. We note that no clear signature of the surface rotation has been detected and in absence of additional constraint, the significant correlation between the splitting and the inclination angle prevents us from constraining them more precisely.

The global parameter of the modes allowed us to have a first estimation of the mass and radius of the star using scaling relations based on solar values: 1.34\,$\pm$\,0.26\,M$_\odot$ and 1.97\,$\pm$\,0.19\,R$_\odot$. A more thorough modeling with AMP, based on the match of the individual mode frequencies and the spectroscopic parameters obtained in this work, provided M\,=\,1.15\,$\pm$\,0.01\,M$_{\odot}$, a radius of R\,=\,1.88\,$\pm$\,0.02\,R$_{\odot}$, and an age $\tau$\,=\,4.33\,$\pm$\,0.12\,Gyr.

\begin{acknowledgements}
The authors thank V. Silva Aguirre, L. Cassagrande and T.~S. Metcalfe for useful comments and discussions. 
This work was partially supported by the NASA grant NNX12AE17G. The CoRoT space mission has been developed and is operated by CNES, with contributions from Austria, Belgium, Brazil, ESA (RSSD and Science Program), Germany and Spain. RAG acknowledges the support given by the French PNPS program. RAG and SM acknowledge the CNES for the support of the CoRoT activities at the SAp, CEA/Saclay. DS acknowledges the support from CNES. SH acknowledges financial support from the Netherlands Organization for Scientific Research (NWO). NCAR is supported by the National Science Foundation. LM, EP, and MR acknowledge financial support from the PRIN-INAF 2010 ({\it Asteroseismology: looking inside the stars with space- and ground-based observations}). KU acknowledges financial support by the Spanish National Plan of R\&D for 2010, project AYA2010-17803. CR, SBF and TRC wish to thank financial support from the Spanish Ministry of Science and Innovation (MICINN) under the grant AYA2010-20982-C02-02. The research leading to 
these results has received funding from the European Community's Seventh 
Framework Programme (FP7/2007-2013) under grant agreement no.~269194 
(IRSES/ASK). Computational time on Kraken at the National Institute of Computational Sciences was provided through NSF TeraGrid allocation TG-AST090107.

\end{acknowledgements}

\bibliographystyle{aa} 
\bibliography{/Users/Savita/Documents/BIBLIO_sav.bib}

\end{document}